\documentclass[12pt]{article}  
\usepackage{graphicx}
\usepackage{epsfig}
\usepackage{hyperref}
\usepackage {amsmath}
\usepackage {amssymb}
\usepackage {bm}  
\usepackage[usenames,dvipsnames]{color}
\usepackage{xspace} 
\newcommand{\pt}{\mbox{$p_T$}\xspace}

\newcommand{\Npart}{\mbox{$N_{\rm part}$}\xspace}

\newcommand{\Nch}{\mbox{$N_{\rm ch}$}\xspace}
\newcommand{\Et}{\mbox{${\rm E}_T$}\xspace}

\newcommand{\sqsn}{\mbox{$\sqrt{s_{_{NN}}}$}\xspace}

\newcommand{\Nqp}{\mbox{$N_{qp}$}\xspace}
\def\lsim{\raise0.3ex\hbox{$<$\kern-0.75em\raise-1.1ex\hbox{$\sim$}}}
\def\gsim{\raise0.3ex\hbox{$>$\kern-0.75em\raise-1.1ex\hbox{$\sim$}}}

\def\mean#1{\left<#1\right>}
\def\Journal#1#2#3#4{{#1}{\bf #2} (#4) #3}
\def\IJMPA{{Int. J. Mod. Phys. A}}
\def\JPG{{J. Phys. G}}
\def\JPCS{{J. Phys: Conf. Series\ }}

\def\NIMA{{Nucl. Instrum. Methods A}}
\def\NPA{{Nucl. Phys. A}}

\def\PLB{{Phys. Lett. B}}
\def\PLC{Phys. Repts.\ }

\def\PRL{Phys. Rev. Lett.\ }
\def\PRD{{Phys. Rev. D}}
\def\PRC{{Phys. Rev. C}}

\def\ARNPS{{Ann. Rev. Nucl. Part. Sci.\ }} 
\def\RMP{Rev. Mod. Phys.\ }

\def\RPP{Rep. Prog. Phys.\ }
\def\QGP{{\color{Red} Q}{\color{Blue} G}{\color{Green} P}} 
\def\QCD{{\color{Red} Q}{\color{Green} C}{\color{Blue} D}}

\normalsize
\begin{document}

\title{Highlights from BNL and RHIC 2015}
\author{M.~J.~Tannenbaum
\thanks{Research supported by U.~S.~Department of Energy, DE-SC0012704.}
\\Physics Department, 510c,\\
Brookhaven National Laboratory,\\
Upton, NY 11973-5000, USA\\
mjt@bnl.gov} 
\date{}
\maketitle
\vspace*{-2pc}
\section{Introduction}\label{sec:introduction}
The Relativistic Heavy Ion Collider (RHIC) at Brookhaven National Laboratory (BNL) is one of the two remaining operating hadron colliders (the other being the LHC at CERN); and the first and only polarized proton collider.  BNL is a multipurpose laboratory, quite different in scope from Fermilab and CERN, with many ``cutting edge'' major research facilities in addition to RHIC. BNL, which is owned by the U.S. Government but operated by a ``management and operating (M\&O) contractor'' was founded by nine major northeastern universities in 1947 to promote basic research in the physical, chemical, biological and engineering aspects of the atomic sciences and for the purpose of the design, construction and operation of large scientific research facilities that individual institutions could not afford to develop on their own. In addition to RHIC, BNL is now home to the National Synchrotron Light-source-II (NSLS-II), the NASA Space Radiation Laboratory, a Tandem Van de Graaf, an Accelerator Test Facility, a Linac Isotope Producer, the Long Island Solar Farm, a Center for Functional Nanomaterials as well as Radiochemistry, Biological Imaging, Environmental \& Climate Sciences groups and a Nonproliferation and National Security Department. BNL has a distinguished history in nuclear \& particle physics and accelerator science~\cite{MJTpasva} as well as many other discoveries~\cite{BNLdiscoveriesWWW}. So far, Nobel Prizes have been awarded to 12 scientists who were either BNL staff members or performed their Nobel work at BNL~\cite{BNLawardsWWW}. 

\section{News from BNL since ISSP2014}
The news this past year was very positive in many ways. The ``management and operating (M\&O) contract'' for BNL, for which there was a solicitation last year~\cite{MJTISSP2014Proceedings}, was awarded to the present management, BSA, which had run the lab for the past 15 years, so that there was a very smooth transition. The major event was the startup and dedication of the ``World's brightest Synchrotoron Light Source'', NSLS-II, by Secretary of Energy Ernest Moniz (who was also an important participant in the USA, France, Germany, UK, China, Russia negotiations with Iran which were ongoing during the ISSP2015 school and reached a historic agreement on a Joint Comprehensive Plan of Action on July 14, 2015 to limit Iran's Nuclear Energy activities to civilian purposes). 

In addition to the successful RHIC run, several other activities took place which related to the future. The proposal for a new ``Large-Acceptance Jet and Upsilon detector for RHIC'' using hadron calorimetry outside a thin-coil superconducting solenoid (originally called sPHENIX) had two important milestones. The laboratory management called a general meeting on June 16, 2015~\cite{sPHENIXworkshop} ``to move expeditiously to form a new detector collaboration to take advantage of the physics opoortunities'' offered by this proposal. The result was the formation of an Institutional Board for the new collaboration, a committee to develop the ByLaws of the new collaboration and to identify spokesperson candidates, working groups to develop a Pre-Conceptual Design Report for a BNL Director's Cost and Schedule review, Nov 9-10, 2015, and a plan to hold a collaboration meeting December 10-12, 2015. 

\begin{figure}[!bh]
\begin{center}
\includegraphics[width=0.20\textwidth]{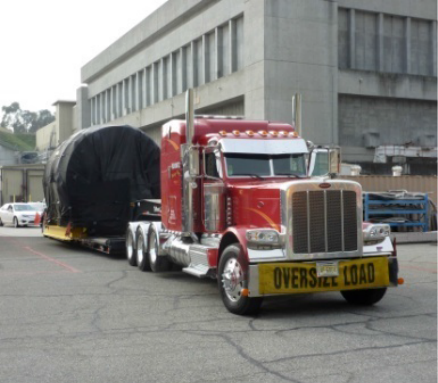}
\includegraphics[width=0.57\textwidth]{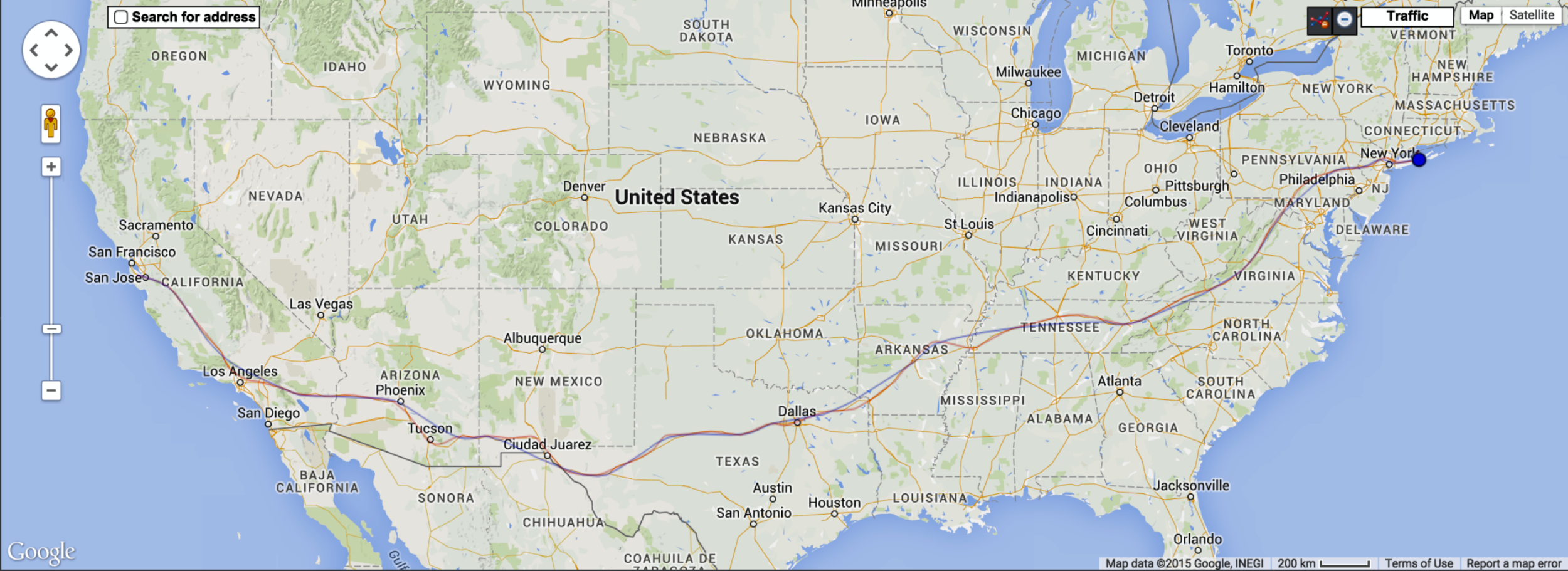}
\includegraphics[width=0.20\textwidth]{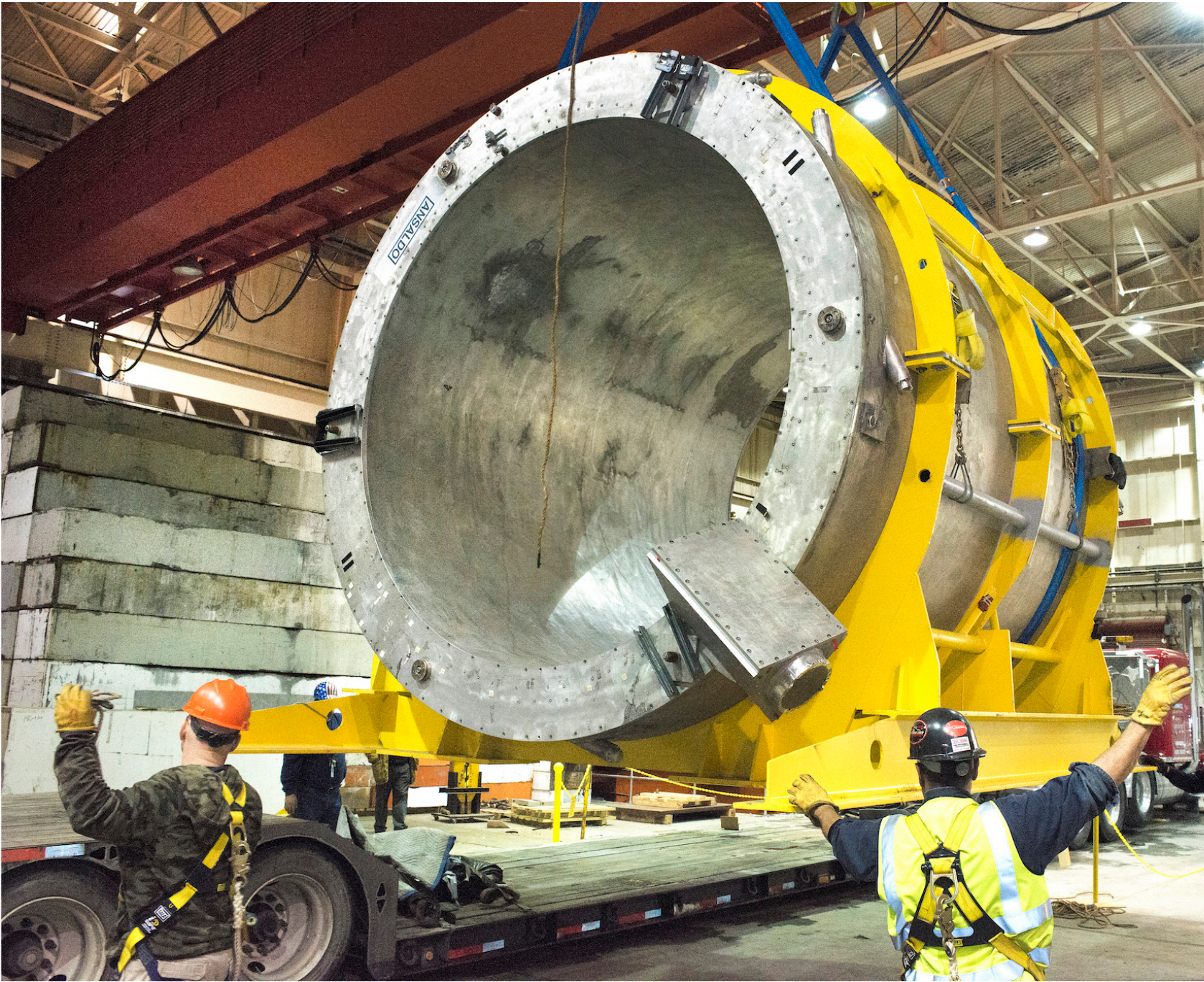}
\end{center}\vspace*{-1.5pc}
\caption[]{\footnotesize a)(left) BABAR solenoid leaves SLAC on Air Ride Trailer b) (center) Route across U.S.A. c) (right) Unloading at BNL.}
\label{fig:BABARmagnetmove}\vspace*{-0.5pc}
\end{figure}

The second milestone was the mid-winter shipment and safe arrival of the (made in Italy) BABAR superconducting solenoid magnet from SLAC on January 16 to BNL on February 3, 2015, which went smoothly (Fig.~\ref{fig:BABARmagnetmove}). The solenoid arrived in mint condition as determined from detailed acceptance tests by the BNL Superconducting Magnet Division.

Also important for the future of Nuclear Physics in the U.S. was the completion of the Long Range Planning exercise for the Nuclear Science Advisory Committee (NSAC) in April 2015, with the final version to be submitted to the Department of Energy (DOE) in October 2015~\cite{NSACLRP2015}, which was unfortunately leaked to the press in May with an article in Nature~\cite{BillionBS} with the headline ``Billion-dollar collider gets thumbs up''. The collider in question is an electron-ion collider, which if located at BNL would be called eRHIC, for which BNL has proposed a highly advanced and energy efficient FFAG electron accelerator/storage ring based on an Energy Recovery Linac that I discussed here last year~\cite{MJTISSP2014Proceedings}. The Nature article made two points that I wish to discuss further. 

\begin{figure}[!t]
\begin{center}
\includegraphics[width=0.35\textwidth]{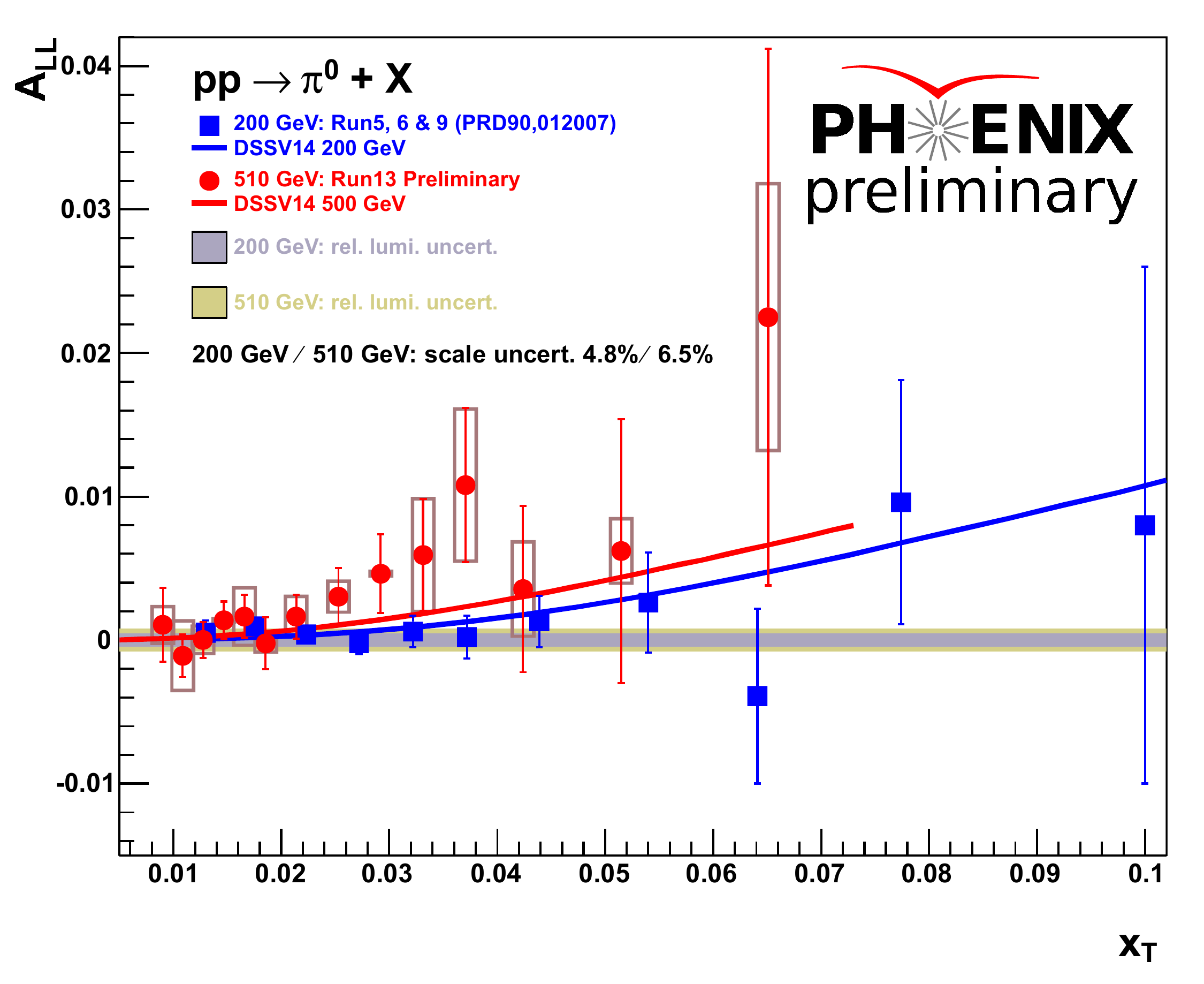}
\includegraphics[width=0.35\textwidth]{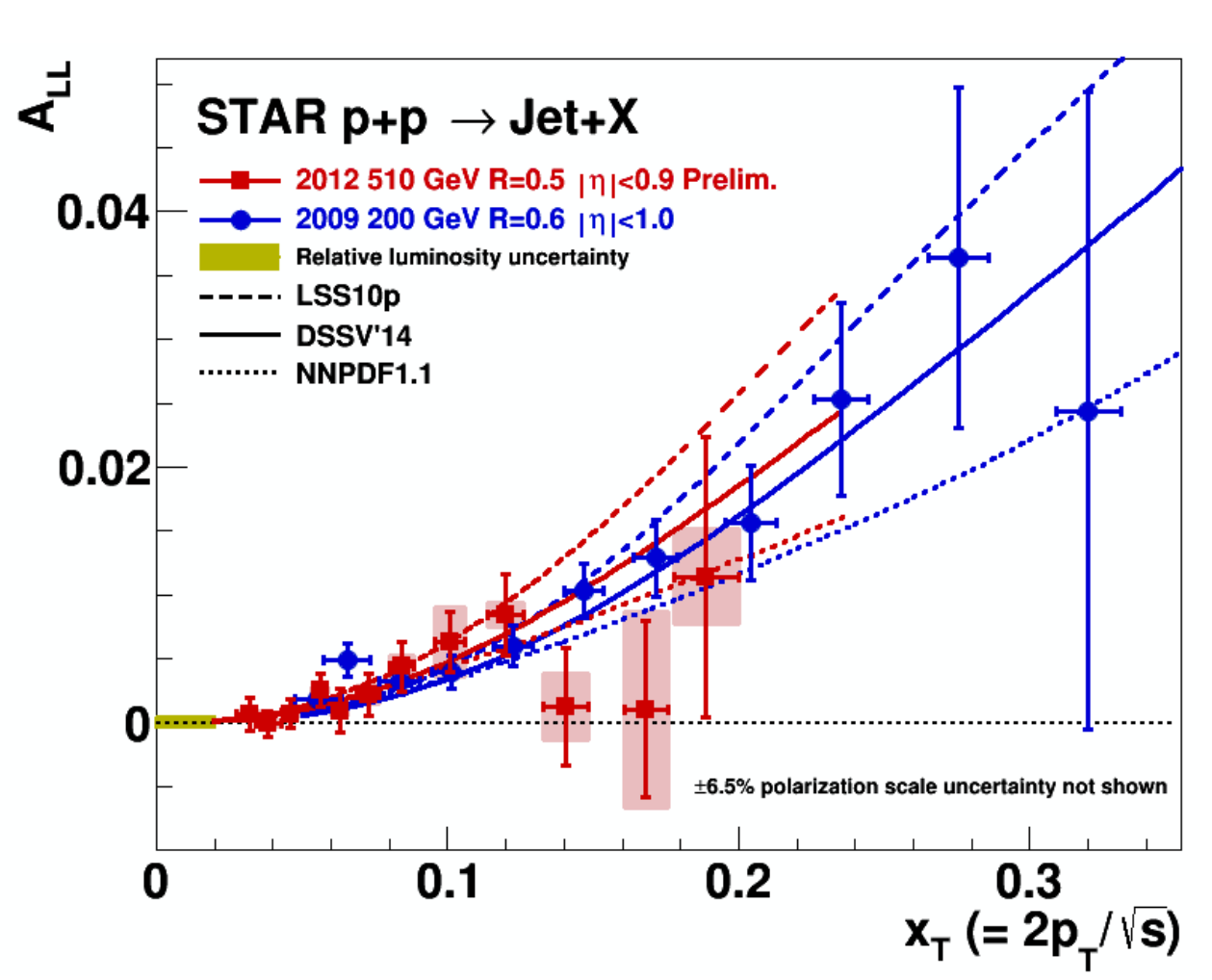}
\includegraphics[width=0.28\textwidth]{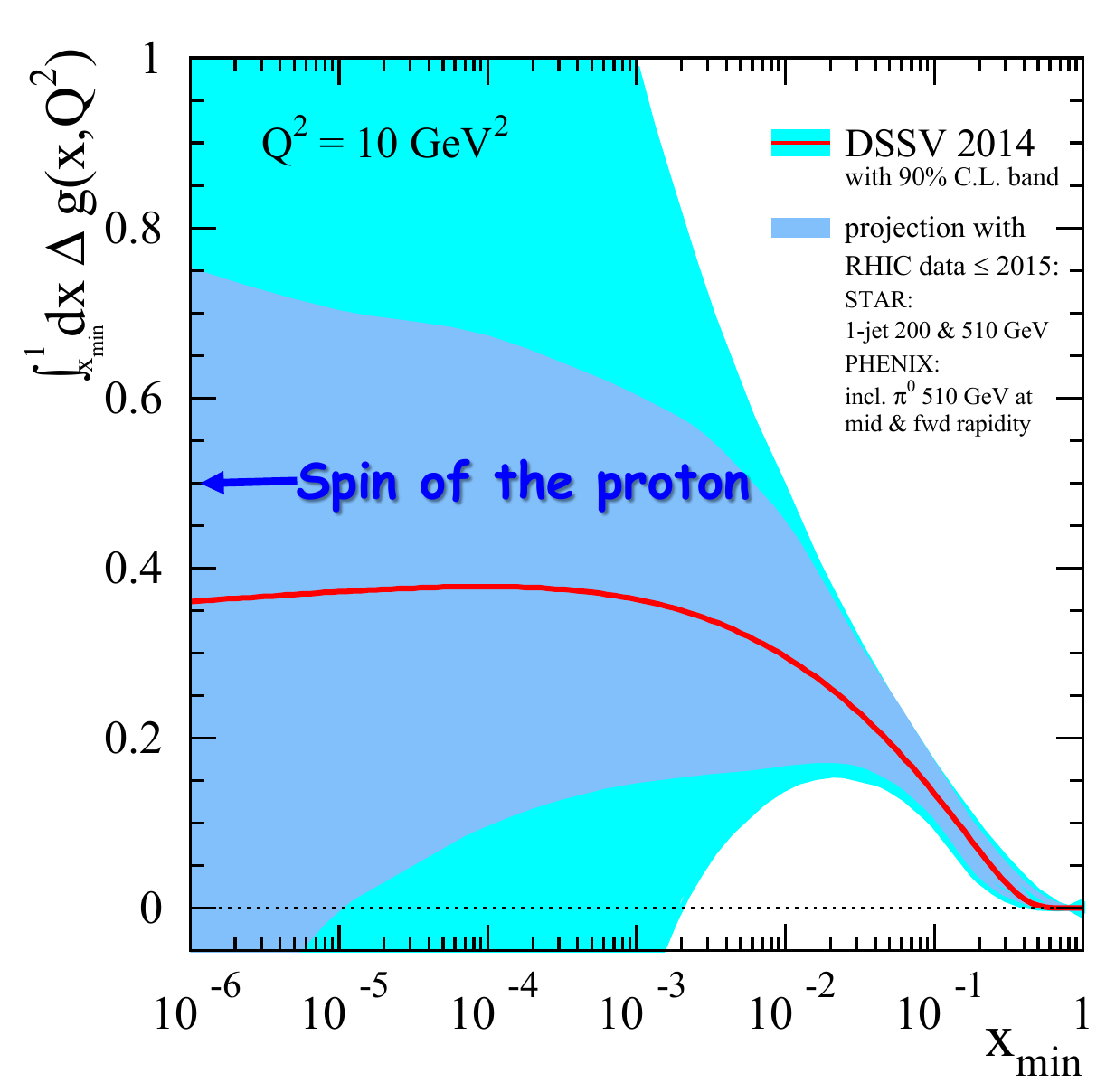}
\end{center}\vspace*{-1.5pc}
\caption[]{\footnotesize a)(left) PHENIX preliminary measurements of $A_{LL}$ of $\pi^0$~\cite{ElkeShura} with DSSV14 fit\cite{DSSV2014} b) (center) STAR measurement of $A_{LL}$ of jets~\cite{ElkeShura} with DSSV14 fit\cite{DSSV2014} c)(right)  Integral of the helicity asymmetry of gluon structure function $\Delta g(x,Q^2)$ from DSSV14 fit\cite{DSSV2014} as a function of $x_{\rm min}$~\cite{ElkeShura}.}
\label{fig:spin}\vspace*{-0.5pc}
\end{figure}

The first involves cost, which a review by an NSAC expert subcommittee~\cite{Temple} put at \$1.5 Billion for either the JLAB or BNL version, which was equal to the JLAB proposal but \$0.5 Billion higher than BNL's proposed cost because of the committee's concern of the higher technical risk of the novel advanced accelerator design. Hopefully, ongoing R\&D and similar work for the CERN LHeC project~\cite{LHeCWorkshop} will overcome the present technical risk and reduce the cost, because many pundits questioned the feasibility of the U.S. alone building an EIC at the \$1.5 Billion price. The second point made in the article was that ``The machine should also solve a puzzle about the proton that has baffled physicists for nearly 30 years''---``strangely the spins of its three constituent quarks add up to only about 1/3 of its own spin.'' Evidently, Nature~\cite{BillionBS} did not know about the latest RHIC-spin results~(Fig.~\ref{fig:spin}) which show a significant non-zero gluon spin contribution to the proton from measurements of $A_{LL}$, the two-spin asymmetry for the scattering of two longitudinally polarized protons:~\cite{BunceVogelsangARNPS}
\begin{equation}
A_{LL}=\frac{1}{P_1 P_2}\frac{(\sigma^{++} +\sigma^{--}) -(\sigma^{+-}+\sigma^{-+})}{(\sigma^{++} +\sigma^{--}) +(\sigma^{+-}+\sigma^{-+})}=\frac{1}{P_1 P_2}\frac{\sigma^{++} -\sigma^{+-}}{\sigma^{++} +\sigma^{+-}}\ \ \mbox{if parity is conserved,}\label{eq:ALL}
\end{equation}
where $\sigma^{++}\equiv N^{++}/L^{++}$ is the measured cross section with both beams having `+' helicity, $N^{++}$ is the measured number of events for an integrated luminosity $L^{++}$, with analogous notation for the other helicity combinations; and $P_1$ and $P_2$ are the polarizations of the two beams. The parton helicity asymmetries are related to the proton asymmetries  by QCD~\cite{BunceVogelsangARNPS}.

For those not familiar with the standard nomenclature:
the the helicity asymmetry  of the structure function $a(x,Q^2)$ for a parton $a$ (where $a$ represents e.g. $u$-quark, $\bar{u}$-quark, gluon, etc.) is defined as $\Delta a(x,Q^2)\equiv a^+(x,Q^2)-a^-(x,Q^2)$ and the ``+'' and ``-'' refer to partons with the same or opposite helicity as the parent proton. There is a helicity sum rule for the spin 1/2 proton:   
$$\frac{1}{2}=\frac{1}{2}\Delta\Sigma + \Delta G +L_q +L_g$$
where $\Delta\Sigma\equiv\int_0^1 dx \left[\Delta u(x,Q^2) +\Delta \bar{u}(x,Q^2) +\Delta d(x,Q^2) +\ldots\right]$ is the combined quark and anti-quark spin contribution, \mbox{$\Delta G\equiv\int_0^1 dx\ \Delta g(x,Q^2)$} is the gluon contribution; and $L_q$ and $L_g$ are possible quark and gluon angular momentum contributions. The quark contribution, which has been measured in DIS~\cite{AidalaRMP} to be $\Delta\Sigma\approx 0.25$, thus accounts for only $\sim 1/4$ of the proton spin. ``The inclusive DIS measurements have, however, very little sensitivity to gluons''~\cite{DSSV2014}. The new results from Fig.~\ref{fig:spin} are the first evidence of a finite gluon polarization in the proton~\cite{DSSV2014}. Notice that the best fit of DSSV14 in Fig.~\ref{fig:spin}c gives $\Delta G\geq 0.36$ for $x_{\rm min}\leq 10^{-3}$.  If improved measurements at $x=10^{-3}$ by future RHIC runs keep the present best fit of DSSV14 with smaller errors, it is conceivable that $\Delta\Sigma/2+\Delta G$ could add up to 1/2 with sufficiently small errors to solve the proton spin puzzle.
\section{RHIC Operations in 2015 and accelerator future plans}\vspace*{-1pc}
\begin{figure}[!t]
\begin{center}
\includegraphics[width=0.325\textwidth]{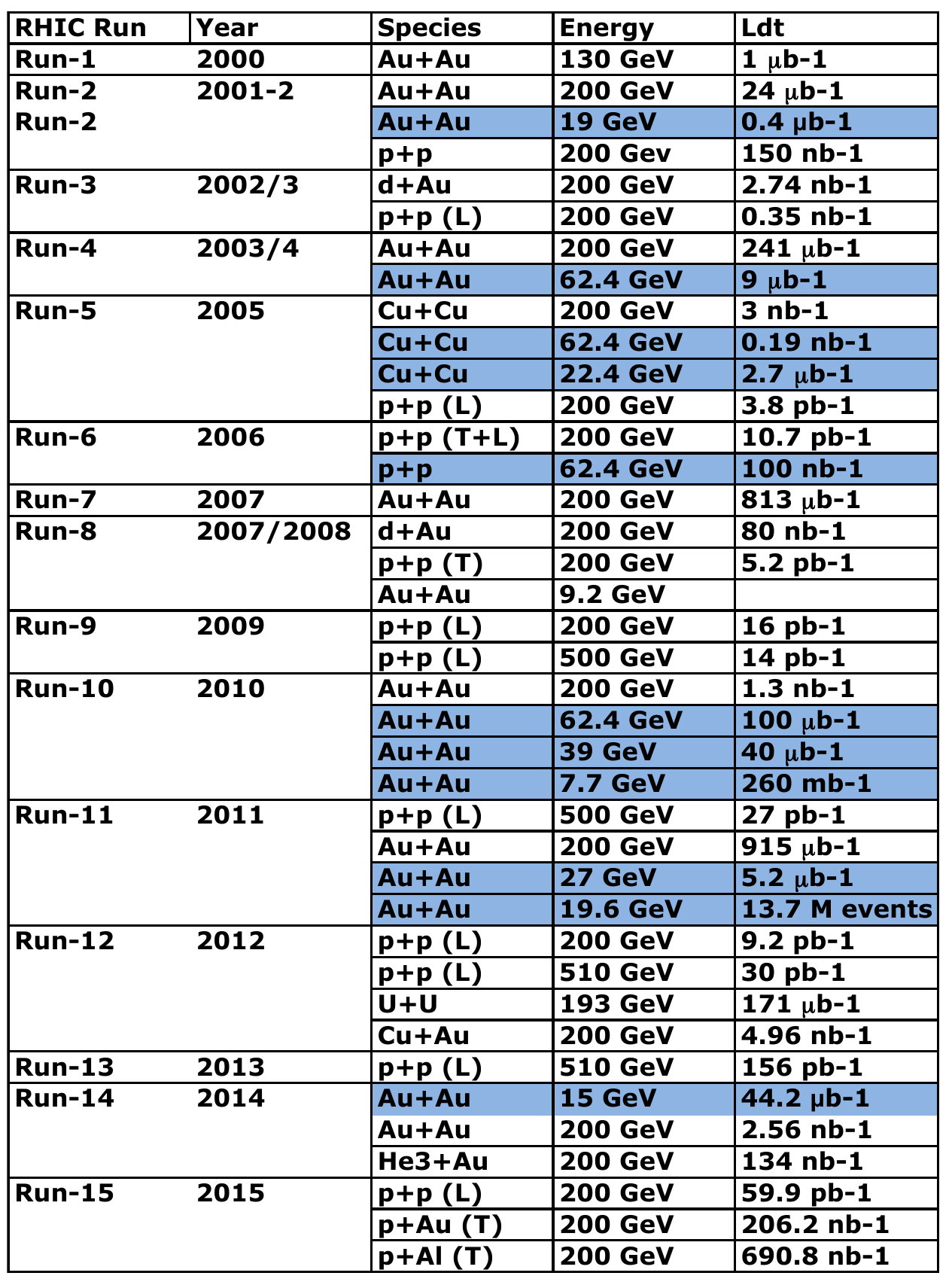}
\includegraphics[width=0.66\textwidth]{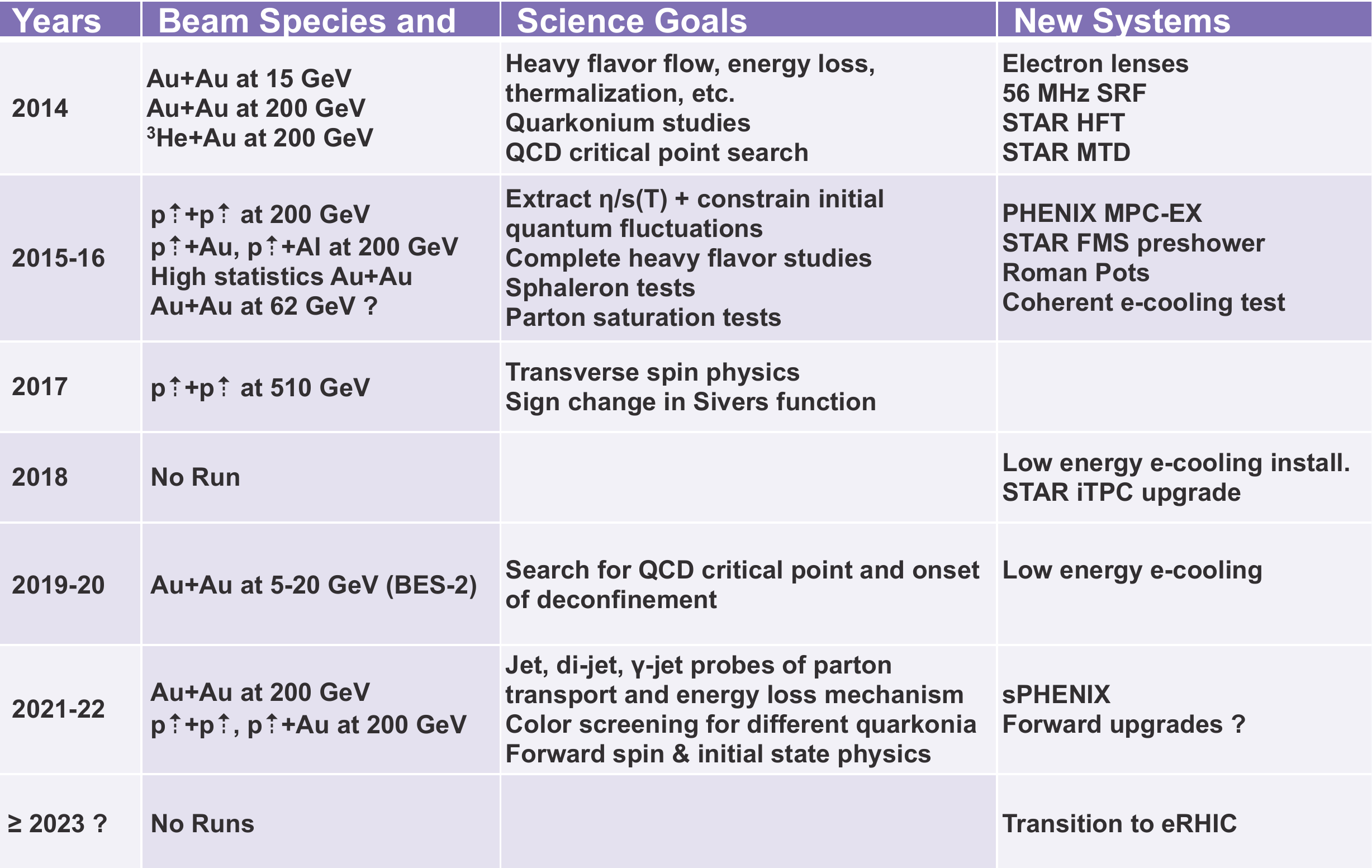}
\end{center}\vspace*{-1.5pc}
\caption[]{\footnotesize a)(left) Year, species and proton polarization ({\bf L}ongitudinal or {\bf T}ransverse), \sqsn and integrated luminosity of RHIC runs. b) (right) Future run schedule and new equipment.}
\label{fig:RHICrunssched}\vspace*{-1.0pc}
\end{figure}
\begin{figure}[!h]
\begin{center}
\includegraphics[width=0.44\textwidth]{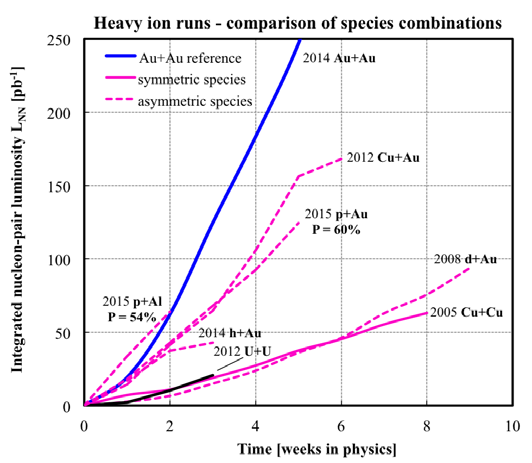}\hspace*{1pc}
\includegraphics[width=0.44\textwidth]{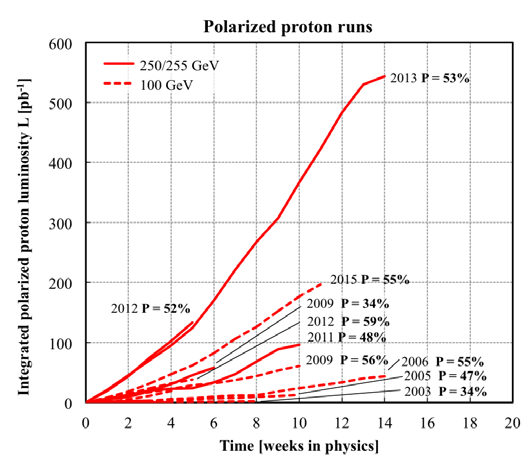}
\end{center}\vspace*{-1.5pc}
\caption[]{\footnotesize a)(left) A$+$B performance, where the nucleon-pair luminosity is defined as $L_{\rm NN}=A\times B\times L$, where $L$ is the luminosity and $A$, $B$ are the number of nucleons in the colliding species. b) (right) Polarized p$+$p performance. Note that the thin lines by the years are pointers. Courtesy Wolfram Fischer.}
\label{fig:RHICperf}\vspace*{-1.0pc}
\end{figure}

Physics data taking for the 15th run at RHIC started on February 11, 2015 and ended on June 22, all at \sqsn=200 GeV. The principal objectives were: i) a comparison p$+$p measurement for the STAR 
 heavy flavor tracker (HFT) which had its first (Au+Au) run in 2014\footnote{The STAR and PHENIX detectors have been operating at RHIC all 15 years~\cite{NIMA499} with regular upgrades.} ; ii) comparison p$+$p measurements for the PHENIX central (VTX) and forward micro-vertex (FVTX) detectors and the new MPC-EX detector (a forward charged particle tracker and EM pre-shower detector to complement the forward EM Calorimeters, located at $3.1\leq|\eta|\leq 3.7$ inside the Muon Pistons); iii) p+Au and p+Al runs for a better baseline of nuclear effects than the previous d+Au  measurements. In all cases the protons were polarized: longitudinally for the p+p  measurements and transversely for the p+A measurements. Fig.~\ref{fig:RHICrunssched}a gives the species, c.m. energies (\sqsn), polarizations and integrated luminosities obtained for all RHIC runs, with new equipment and future plans given in Fig.~\ref{fig:RHICrunssched}b. 
The luminosity performance is given in Fig.~\ref{fig:RHICperf}. Note that the p$+$p integrated luminosity at \sqsn=200 GeV in 2015 exceeds the sum of all previous p+p integrated luminosity at this \sqsn. \vspace*{-2.0pc}
\subsection{Nobody's perfect---the first asymmetric Z/A run. }
\begin{figure}[!h]
\begin{center}
\includegraphics[width=0.88\textwidth]{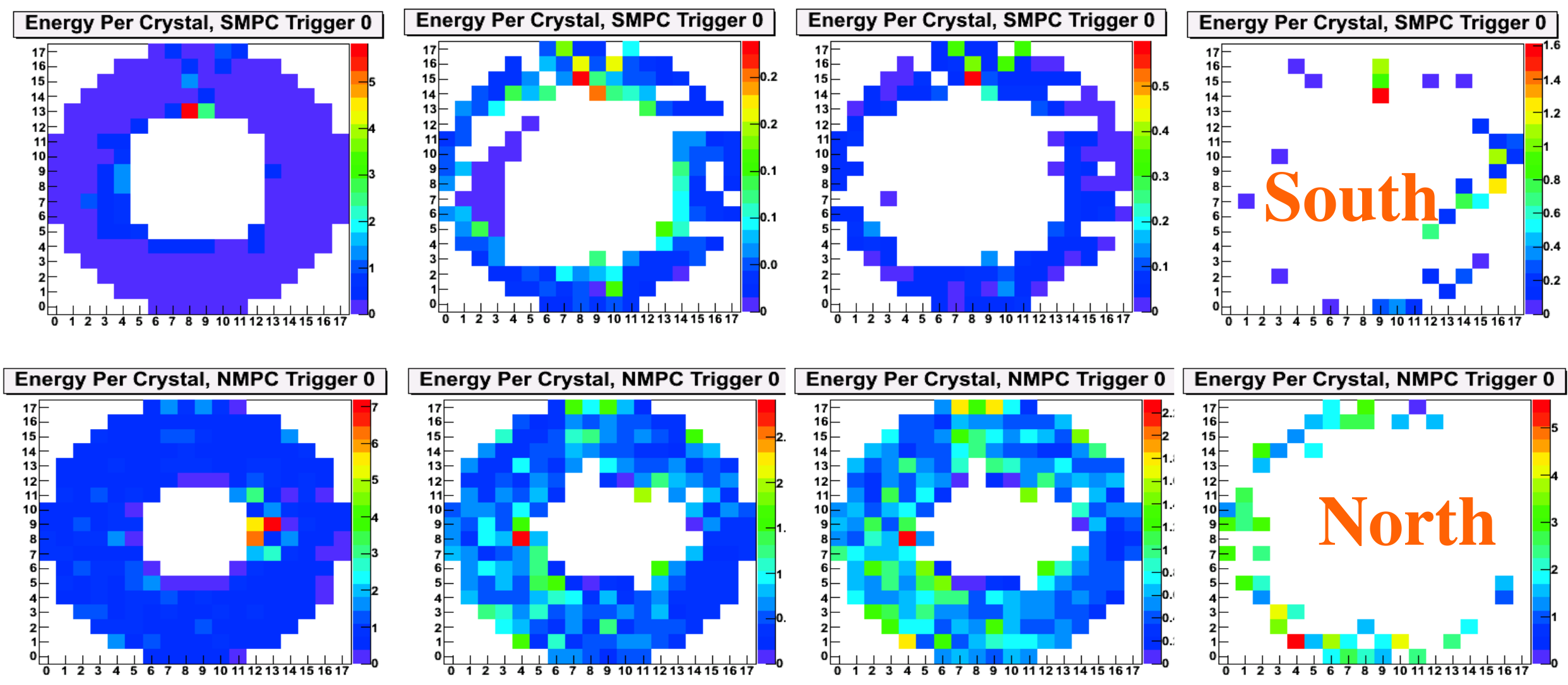}
\end{center}\vspace*{-1.5pc}
\caption[]{\footnotesize MPC damage from yellow-ring abort kicker prefires: top row South MPC, bottom row North MPC. From left to right: Start of run, Feb. 11; after May 11; after May 28; after June 1.  }
\label{fig:MPCfades}\vspace*{-0.5pc}
\end{figure}

Because of the asymmetric p+A running, the final focussing DX magnets near the interaction regions had to be moved. Aperture restrictions were identified and fixed; masks for the protection of the detectors in case of a quench were installed. However since this was the first such run at RHIC, abort kicker prefires were not adequately shielded and the MPC detector was damaged by several prefires and removed from service on June 1 (Fig.~\ref{fig:MPCfades}). The VTX and FVTX detectors were also removed as a precaution. The good news is that only the p$+$Al measurement was affected by these problems.

\section{10th Anniversary Celebration of the Perfect Liquid}
High energy nucleus-nucleus collisions provide the means of creating nuclear matter in conditions of extreme temperature and density where a phase transition is expected from a state of nucleons containing confined quarks and gluons to a state of quarks and gluons, deconfined from their original nucleons, covering a volume that is many units of the confinement length scale ($\sim 1$ fm) in which the $q$ and $g$, with their color charge fully exposed, freely traverse the medium composed of a large density of similarly exposed color charges. This state of nuclear matter was originally given the name Quark Gluon Plasma (\QGP)~\cite{Shuryak80}, a plasma being an ionized gas. 
The \QGP\ was discovered at RHIC, and announced on April 19, 2005. However the results at RHIC~\cite{seeMJTROP} indicated that instead of behaving like the anticipated gas of free quarks and gluons, the matter created in heavy ion collisions at nucleon-nucleon c.m. energy $\sqrt{s_{NN}}=200$ GeV appears to be more like a {\em liquid}. This matter interacts much more strongly than originally expected, as elaborated in peer reviewed articles by the 4 RHIC experiments~\cite{BRWP,PHWP,STWP,PXWP}, which inspired the theorists~\cite{THWPS} to give it the new name ``s\QGP" (strongly interacting \QGP). These properties were quite different from the ``new state of matter'' claimed in a press-conference~\cite{CERNBaloney} by the CERN fixed target heavy ion program on February 10, 2000, which was neither peer-reviewed nor published. 

I had a few additional quips in my discussion of this discovery in previous ISSP talks and proceedings~\cite{MJTIJMPA2014}. Perhaps I was too harsh because in an older previous publication~\cite{seeMJTROP} I noted that it was not unreasonable to expect surprises in the search for the \QGP, i.e. ``Many of the predicted properties will be found but will not be the \QGP; the \QGP\ will have a few unpredicted or unexpected properties; the search will uncover many unexpected backgrounds and new properties of p$+$p and A$+$A collisions, some of which may be very interesting phenomena in their own right.'' In fact this is what has happened\footnote{It may be interesting for some readers to look back at some of the early predictions for the \QGP~\cite{MJToldies}.} and is still happening this year. 

At the time of the proposals for experiments at RHIC (also for the LHC~\cite{ALICEJINST}) in 1990-91, the ``gold-plated''signature for  deconfinement in the \QGP\ was the suppression of $J/\Psi$ production caused by the Debye screening of the \QCD\  potential in the \QGP\ so that the $c$ and $\bar{c}$ quarks could not bind~\cite{MatsuiSatz86}. However, as predicted a year later, in 1987~\cite{MatsuiLBL24604}, enhanced production of $c$ and $\bar{c}$ quarks in A$+$A collisions and recombination of $c\bar{c}$ into $J/\Psi$ might (and did~\cite{MJTIJMPA2014}) ``hinder'' $J/\Psi$ suppression as evidence for the \QGP. 

In 1997, an additional signal of \QGP\ formation was proposed~\cite{BDMPS}, Jet Quenching---energy loss from coherent LPM radiation of hard-scattered partons exiting the \QGP, which would result in an attenuation of the jet energy and broadening (enhanced acoplanarity) of the di-jets formed by fragmentation of the outgoing partons. This has been a robust signature of \QGP\ formation (although more complicated than the original proposal) and is one of the few remaining signals without an alternate explanation. 

In my opinion, the key \QGP\ discoveries at RHIC, so far, are the following:\vspace*{-0.8pc}
\begin{itemize}
\item Suppression of high $p_T$ hadrons from hard-scattering of initial state partons~\cite{ppg003}; also modification of the away-side jet~\cite{HardkeNPA715}.\vspace*{-0.8pc}
\item Elliptic flow at the Hydrodynamic limit as a near ideal fluid~\cite{TeaneyPRC68} with shear viscosity/entropy density, $\eta/s$, at or near the quantum lower bound $\eta/s\approx \frac{1}{4\pi}$~\cite{Kovtun05}.\vspace*{-0.8pc}
\item Elliptic flow of particles proportional to the number of constituent quarks~\cite{seeVoloshin2005,PXv2PIDPRL91,PXPRL98}.\vspace*{-0.8pc}
\item Mid-rapidity transverse energy, $d\Et/d\eta$ and charged particle multiplicity density $d\Nch/d\eta$ proportional to the number of constituent quark participants \Nqp.~\cite{EreminVoloshin,ppg100}.\vspace*{-0.8pc}
\item Higher order flow moments proportional to density fluctuations of the initial colliding nuclei~\cite{AlverRoland,ppg132}.\vspace*{-0.8pc}
\item Suppression and flow of heavy quarks roughly the same as that of light quarks~\cite{ppg066}.\vspace*{-0.8pc} 
\item \QCD\ hard prompt photons not suppressed~\cite{ppg042}, do not  flow~\cite{ppg126}.\vspace*{-0.8pc}
\item Production~\cite{ppg086} and flow~\cite{ppg126} of thermal soft photons.  
\end{itemize}\vspace*{-0.8pc}

\subsection{A few illustrations}
\subsubsection{$d\Et/d\eta$ and  $d\Nch/d\eta$ proportional to constituent quark participants \Nqp.}
\begin{figure}[!h]\vspace*{-0.5pc} 
      \centering
\raisebox{0pc}{\includegraphics[width=0.30\linewidth]{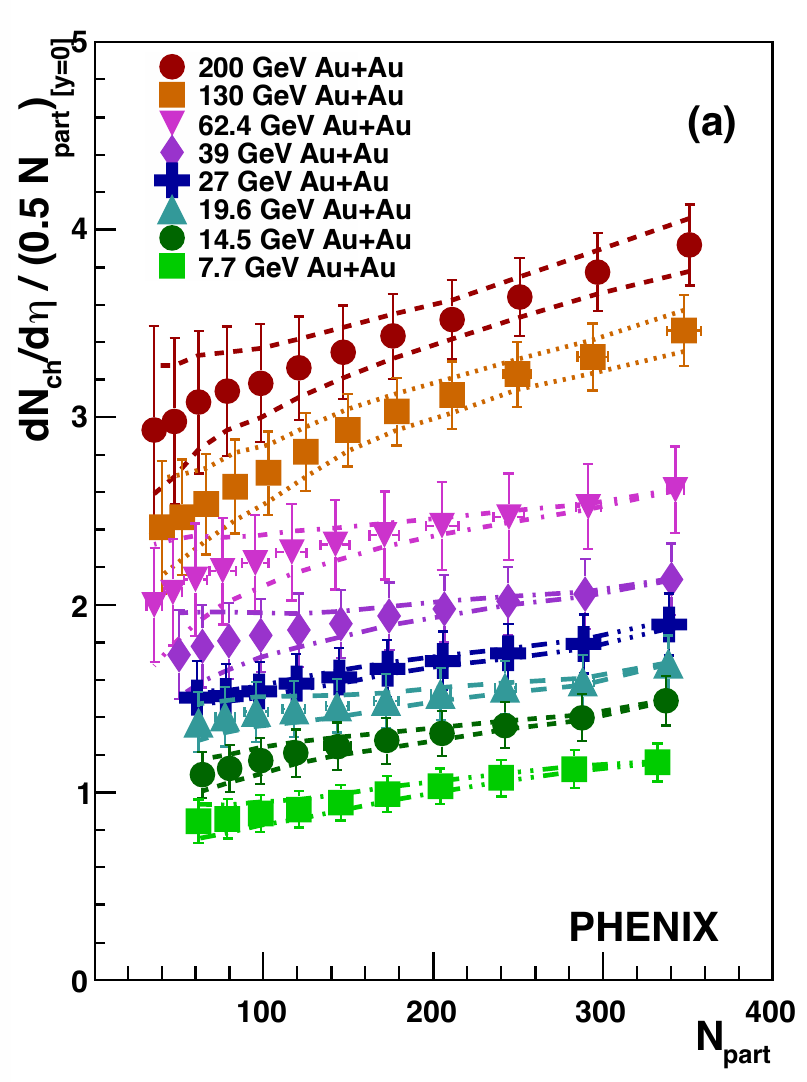}}\hspace*{0.4pc} 
\raisebox{0pc}{\includegraphics[width=0.30\linewidth,angle=0]{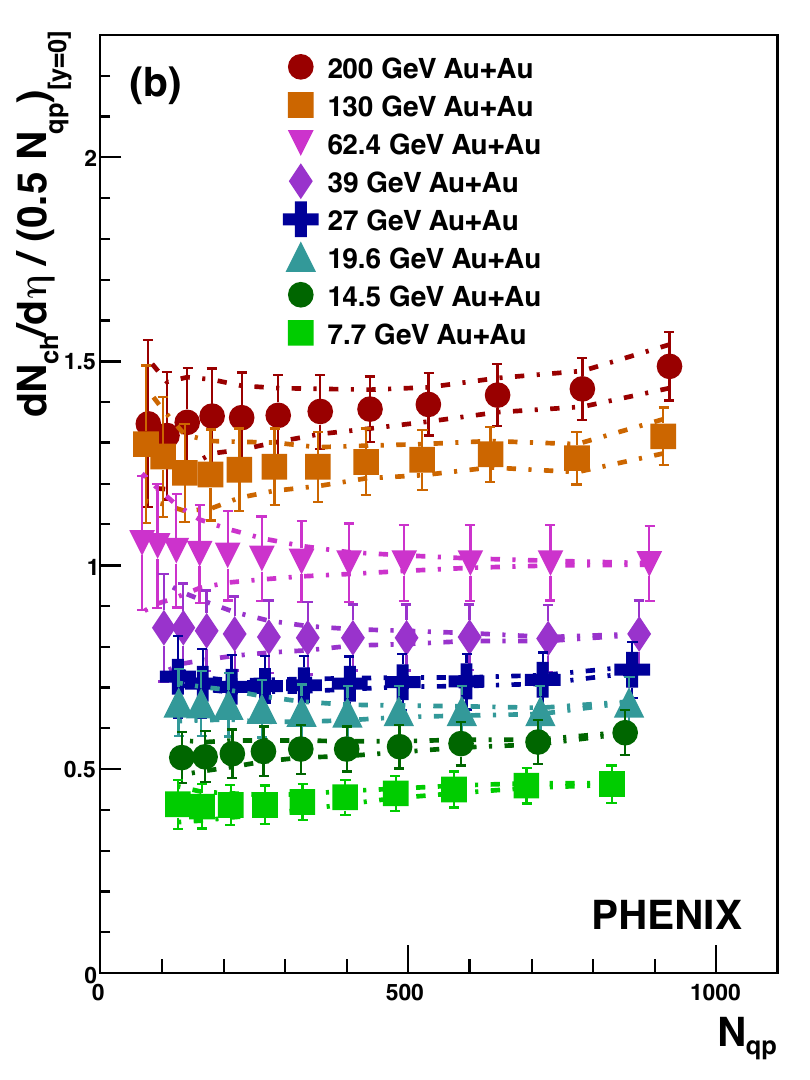}}\hspace*{0.5pc}
\raisebox{0pc}{\includegraphics[width=0.30\linewidth,angle=0]{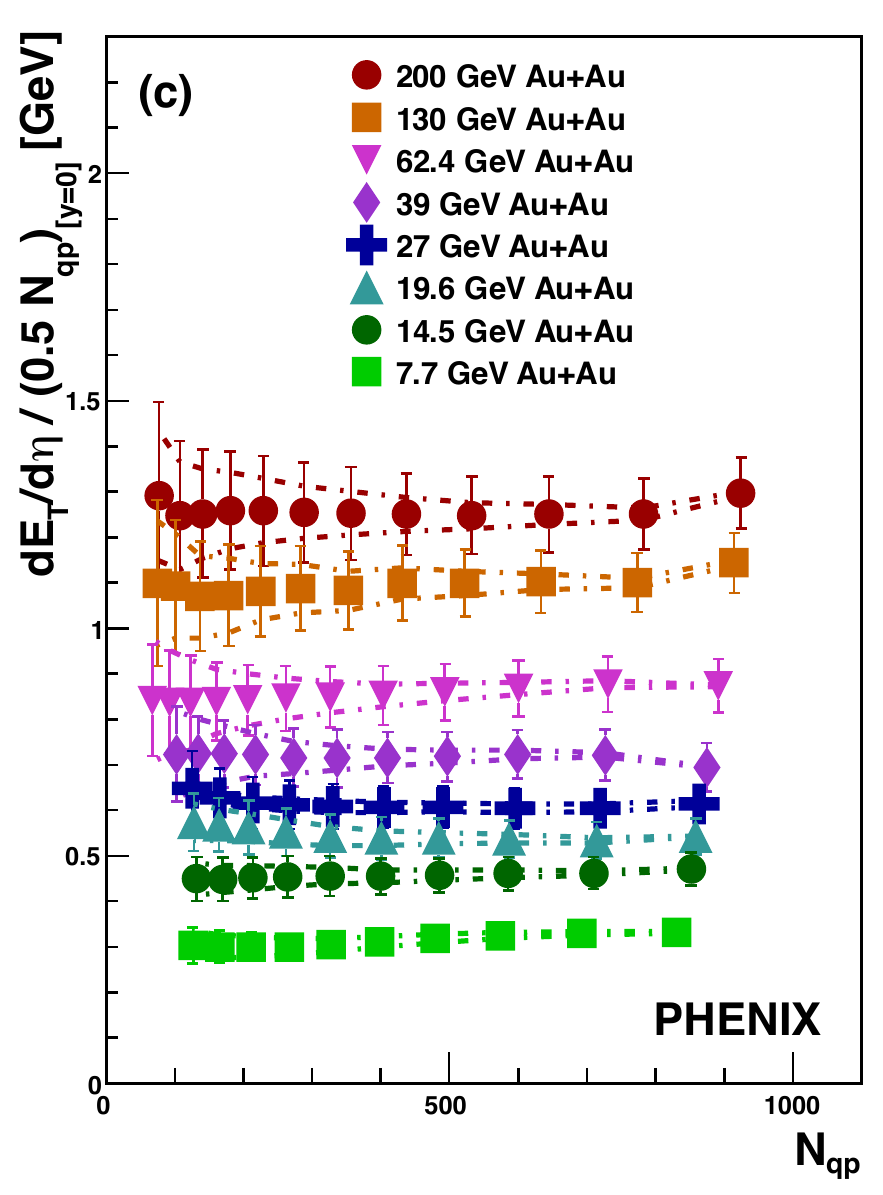}}
\normalsize
     \caption[]{\footnotesize PHENIX\cite{ppg174} mid-rapidity (a) $d\Nch/d\eta/(0.5\Npart)$ as a function of \Npart,   (b) $d\Nch/d\eta/(0.5\Nqp)$ (c) $d\Et/d\eta/(0.5\Nqp)$ as a function of \Nqp, in Au$+$Au collisions at the \sqsn\ indicated.}
      \label{fig:NqpvsNpart}
   \end{figure}

The closely related charged particle density $d\Nch/d\eta$ and transverse energy density $d\Et/d\eta$ at mid-rapidity are composed of soft particles with low transverse momenta and a steeply falling exponential spectrum, $d\sigma/\pt d\pt \propto \exp{-6\pt}$ in p$+$p collisions. The first measurements~\cite{HalliwellPRL39} of the total charged multiplicity \Nch in p$+$A collisions at nucleon-nucleon c.m. energies \sqsn=10-20 GeV showed that \Nch was proportional to the number of nucleons, \Npart, that participate (i.e are struck) in the collision of the nuclei.~\footnote{\Npart is an estimate of the centrality of the collision, the overlap of the two colliding nuclei~\cite{seeMJTROP}.} \Npart\ scaling was originally expected to be valid at RHIC energies but was not observed at mid-rapidity (Fig.~\ref{fig:NqpvsNpart}a). The correct scaling found at mid-rapidity (including \sqsn=10-20 GeV) is the number of (massive) constituent quark participants \Nqp ~\cite{ppg174} (Fig.~\ref{fig:NqpvsNpart}b,c).
\vspace*{-1.0pc}
\subsubsection{Collective Flow---anisotropic production of particles.}
\begin{figure}[!t]
   \begin{center}
\raisebox{1.0pc}{\includegraphics[width=0.45\linewidth]{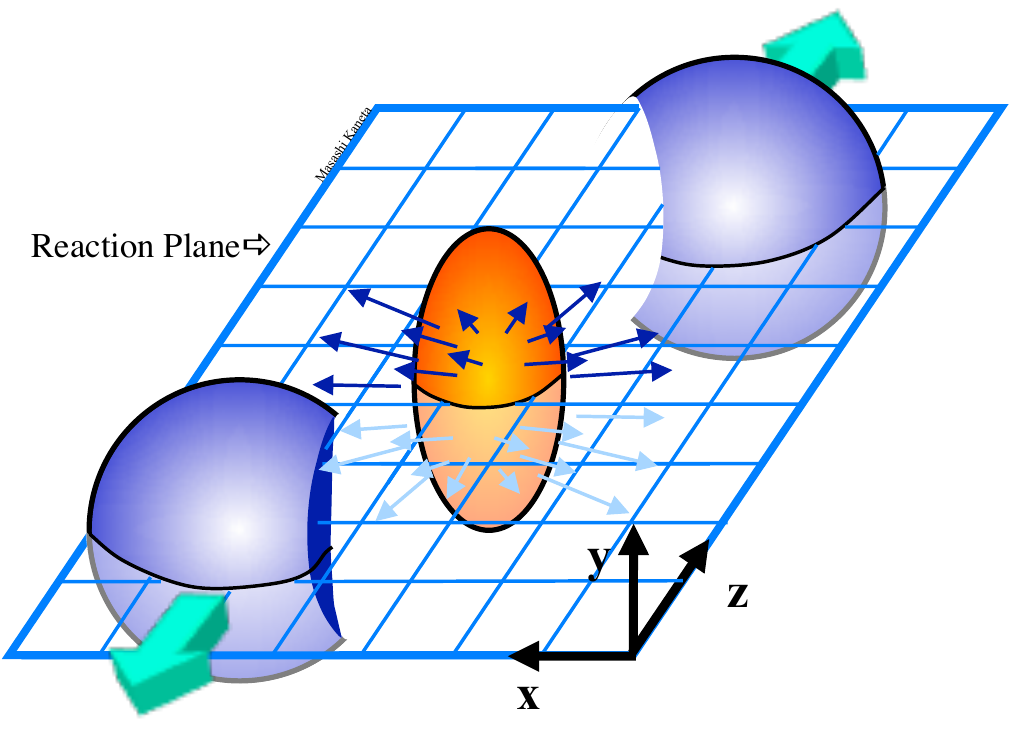}}\hspace*{2.2pc}
\raisebox{0.0pc}{\includegraphics[width=0.27\linewidth]{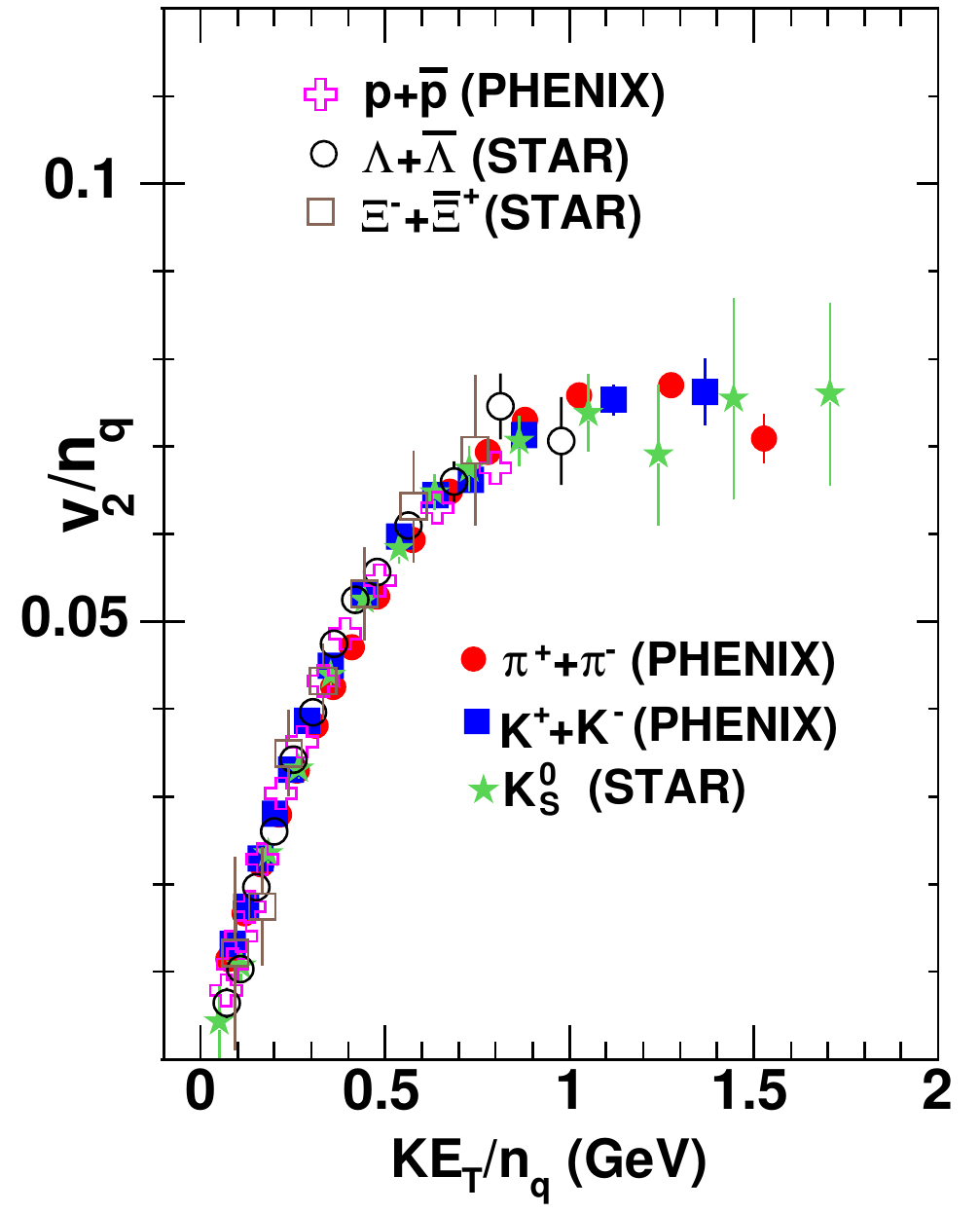}}
\end{center}\vspace*{-2.0pc}
\caption[]{\footnotesize (a) (left) Almond shaped overlap zone generated just after an A$+$A collision where the incident nuclei are moving along the $\pm z$ axis. The reaction plane by definition contains the impact parameter vector (along the $x$ axis)~\cite{KanetaQM04}. (b) (right) $v_2/n_q$ vs. $KE_T/n_q$ for baryons and mesons~\cite{PXv2PRL98}.  
\label{fig:MasashiFlow}}\vspace*{-1.0pc}
\end{figure}

Collective flow (anisotropic production of particles)~\cite{Ollitrault}, or simply flow, is a collective effect which can not be obtained from a superposition of independent N$+$N collisions. ``It does not assume necessarily the hydrodynamic flow, which in particular would require a thermalization of the system''~\cite{seeVoloshin2005} but this distinction is not generally discussed. The almond shaped overlap region of the A+A collision causes the particles to be emitted more favorably in the reaction plane (see Fig.~\ref{fig:MasashiFlow}a).  
The semi-inclusive single particle spectrum is modified by an expansion in harmonics of the azimuthal angle of the particle with respect to the reaction plane, $\phi-\Phi_R$:\vspace*{-0.8pc}   
\begin{equation}
{Ed^3 N \over dp^3}={d^3 N\over p_T dp_T dy d\phi}
={d^3 N\over 2\pi\, p_T dp_T dy} \left[ 1+\sum_n 2 v_n \cos n(\phi-\Phi_R)\right] .
\label{eq:siginv2}
\end{equation} 
The Fourier coefficient $v_2$, called elliptic flow, is predominant at mid-rapidity. Odd harmonics were thought to be forbidden by the symmetry $\phi\rightarrow \phi+\pi$ of the almond. Only in 2010 was it realized that the nuclear geometry  fluctuated from event to event and did not respect the average symmetry~\cite{AlverRoland}. 

Flow measurements contributed two of the most important results about the properties of the \QGP: i) the scaling of $v_2$ of identified particles at mid-rapidity with the number of constituent-quarks $n_q$ in the particle---$v_2/n_q$ scales with the transverse kinetic energy per constituent-quark, $KE_T/n_q$, because particles have not formed at the time flow develops (Fig.~\ref{fig:MasashiFlow}b); ii) the  persistence of flow for $p_T>1$ GeV/c which implied that the viscosity is small~\cite{TeaneyPRC68}, perhaps as small as a quantum viscosity bound from string theory~\cite{Kovtun05}, $\eta/s=1/(4\pi)$, where $\eta$ is the shear viscosity and $s$ the entropy density per unit volume.  This led to the description of the ``s\QGP'' produced at RHIC as ``the perfect fluid''. 
\subsubsection{More surprises this year, flow in three small systems.}
\begin{figure}[!h]
\begin{center}
\includegraphics[width=0.32\textwidth]{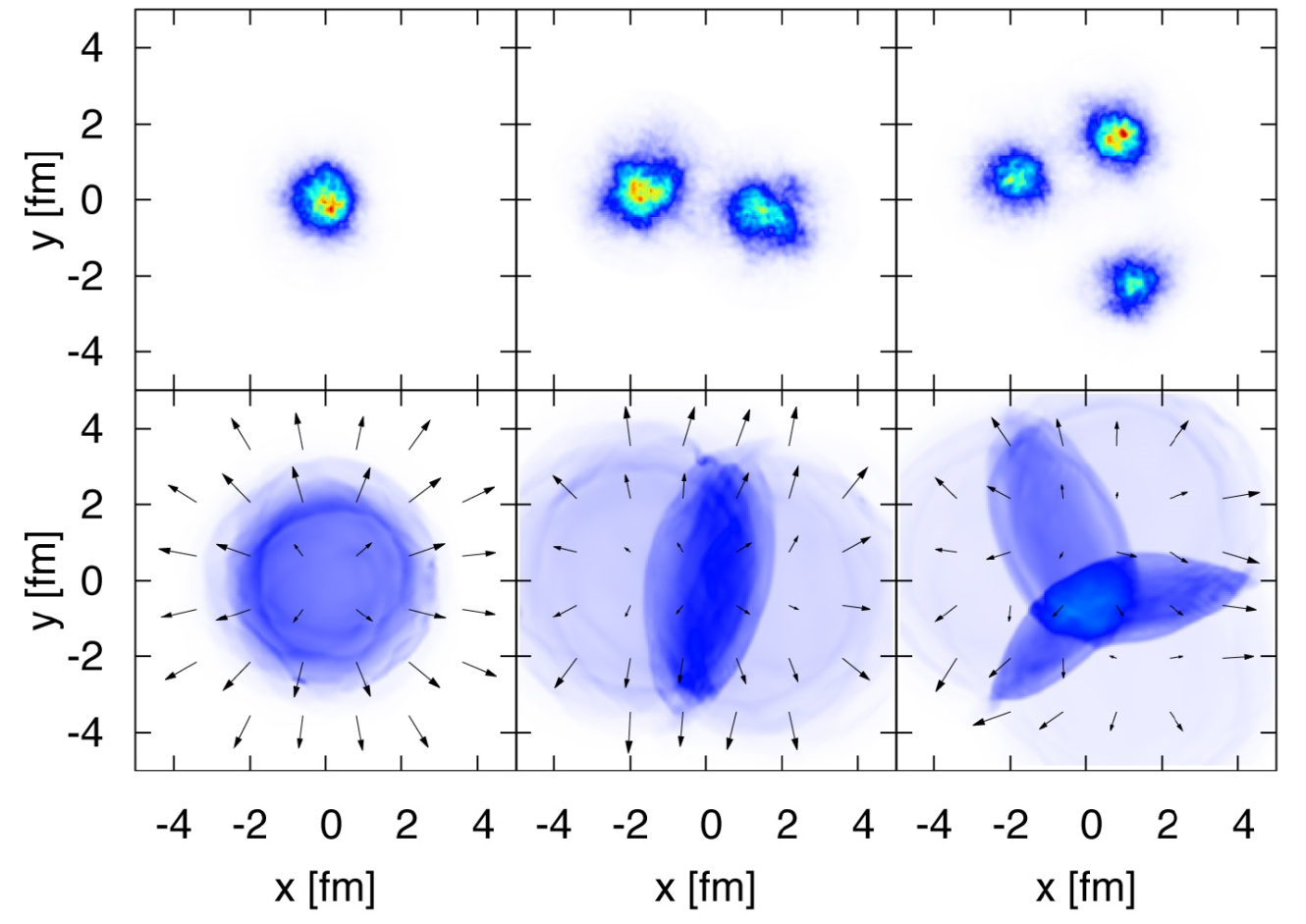}
\includegraphics[width=0.32\textwidth]{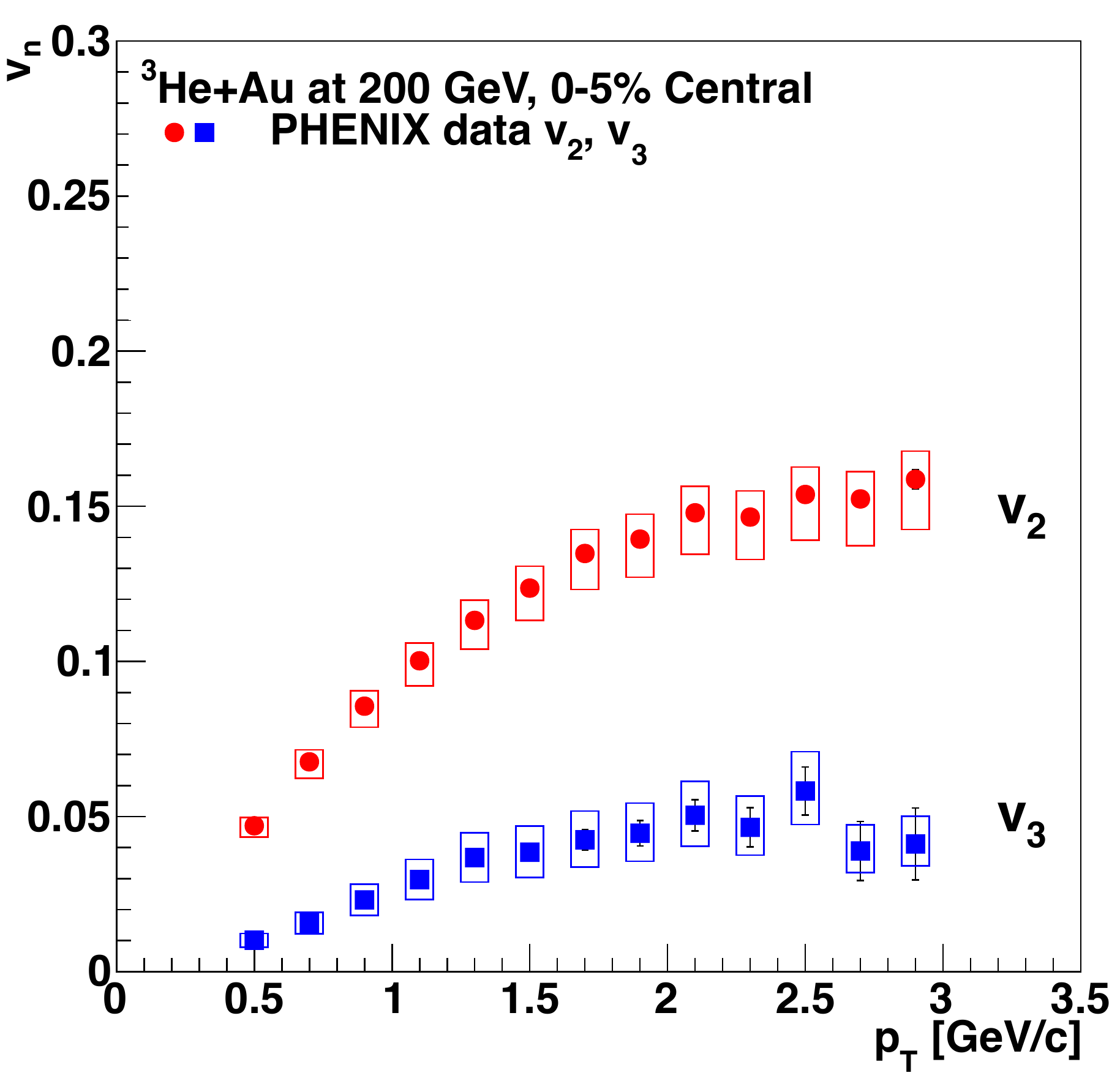}
\includegraphics[width=0.32\textwidth]{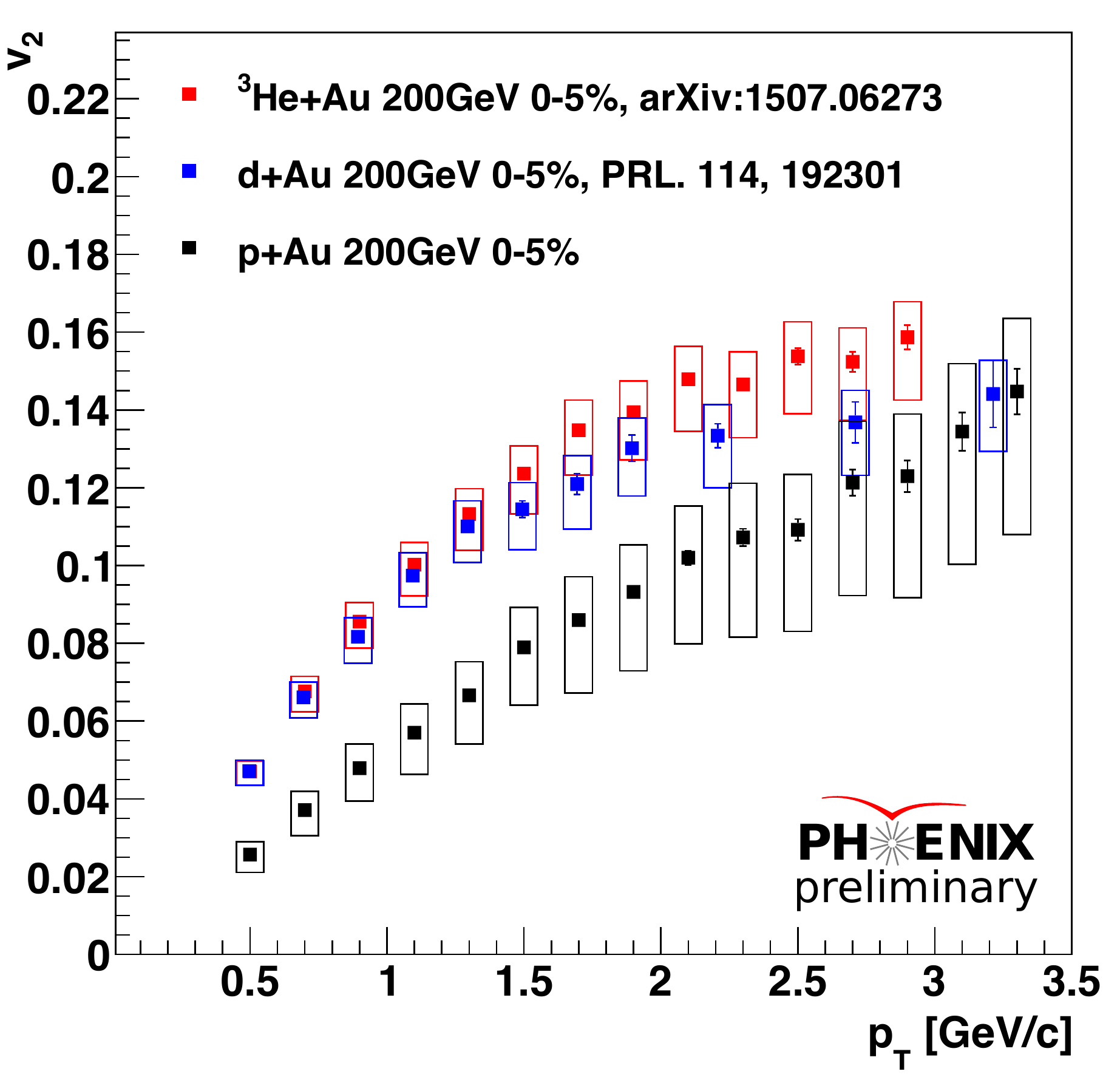}
\end{center}\vspace*{-1.5pc}
\caption[]{\footnotesize a) (left) (above) Simulated initial energy depositions for p, d and He$^3$ on a larger nucleus; (below) the \QGP\ geometry resulting from hydrodynamic flow~\cite{NSACLRP2015}. b) (center) PHENIX~\cite{PXHe3PRL115}  $v_2$ and $v_3$ from He${^3}+$Au with 0-5\% centrality at \sqsn=200 GeV. c)(right) $v_2$ from p,d,He$^3 +$Au at \sqsn=200 GeV\cite{ItaruQM2015}. }
\label{fig:pdHe3}\vspace*{-0.5pc}
\end{figure}

Last year~\cite{MJTISSP2014Proceedings}, I noted that results in 2013 observed what looked very much like collective flow in p+Pb at LHC and d+Au at RHIC. These systems were believed to be too small for collective effects which inspired the He${^3}+$Au run at RHIC in 2014 to see whether triangular flow ($v_3$) would be more prominent with a 3 nucleon projectile (Fig.~\ref{fig:pdHe3}a). 
The He${^3}+$Au result (Fig.~\ref{fig:pdHe3}b)~\cite{PXHe3PRL115} did indeed show a strong $v_3$ as well a $v_2$ comparable to the d$+$Au (Fig.~\ref{fig:pdHe3}c). Perhaps no longer surprising, the p$+$Au data from this year's run~\cite{ItaruQM2015} also showed a comparable $v_2$, but smaller, consistent with the more spherical initial geometry. Of special interest is how well the large $v_3/v_2$ ratio in He$^3+$Au corresponds to the ratio of the eccentricities in the initial geometry because translation of the geometrical anistropy into the flow anisotropy depends on the viscosity and dissipates with larger viscosities for larger $v_n$. The issue of whether this is really hydrodynamic flow is still one of the very hot topics in RHI physics. \vspace*{-1.0pc}
\subsubsection{Jet quenching at RHIC --- Suppression of high $p_T$ particles }
\begin{figure}[!h]
   \begin{center}
\raisebox{0.0pc}{\includegraphics[width=0.384\textwidth]{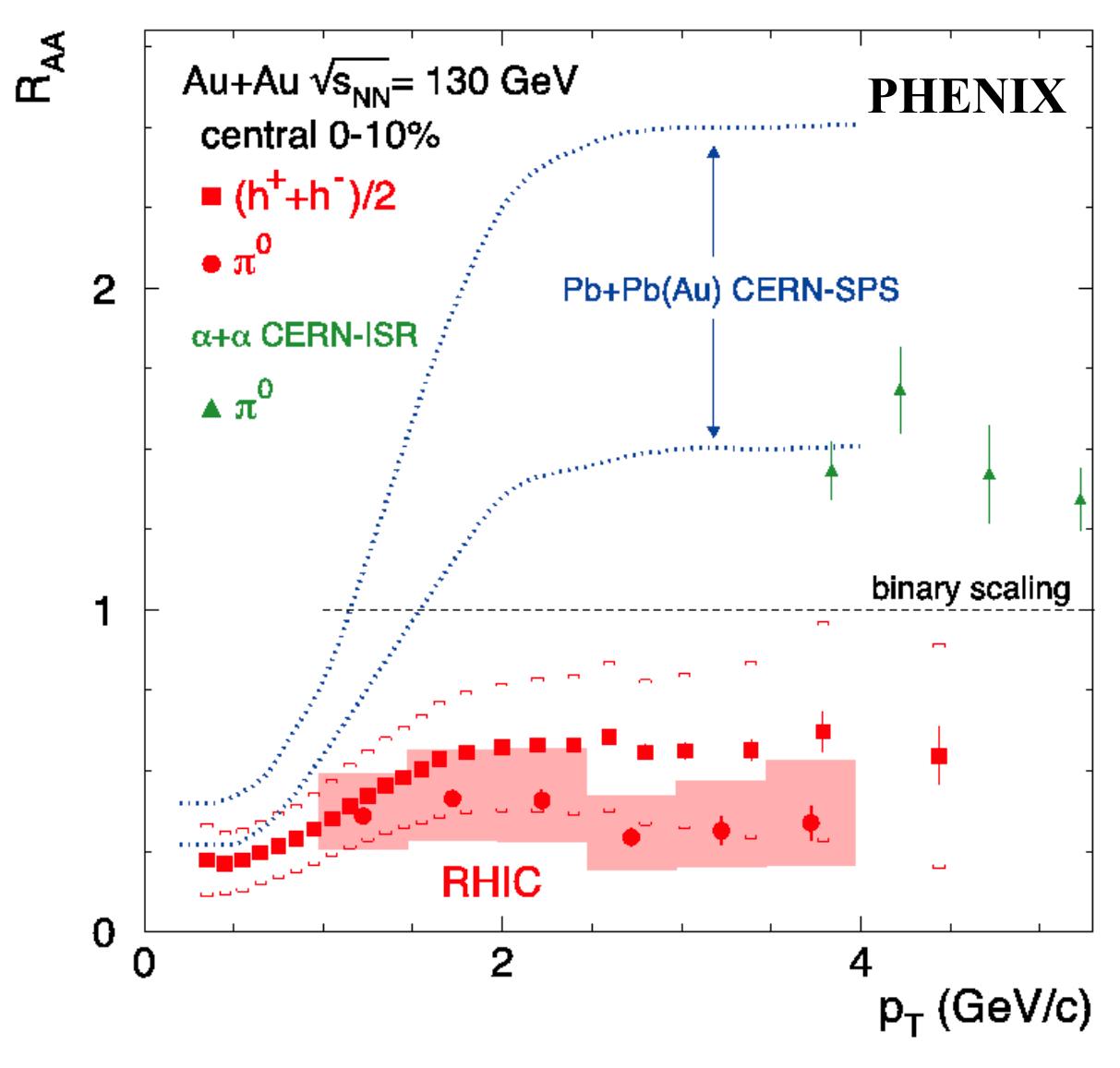}}\hspace{3pc}
\raisebox{0.8pc}{\includegraphics[width=0.36\textwidth]{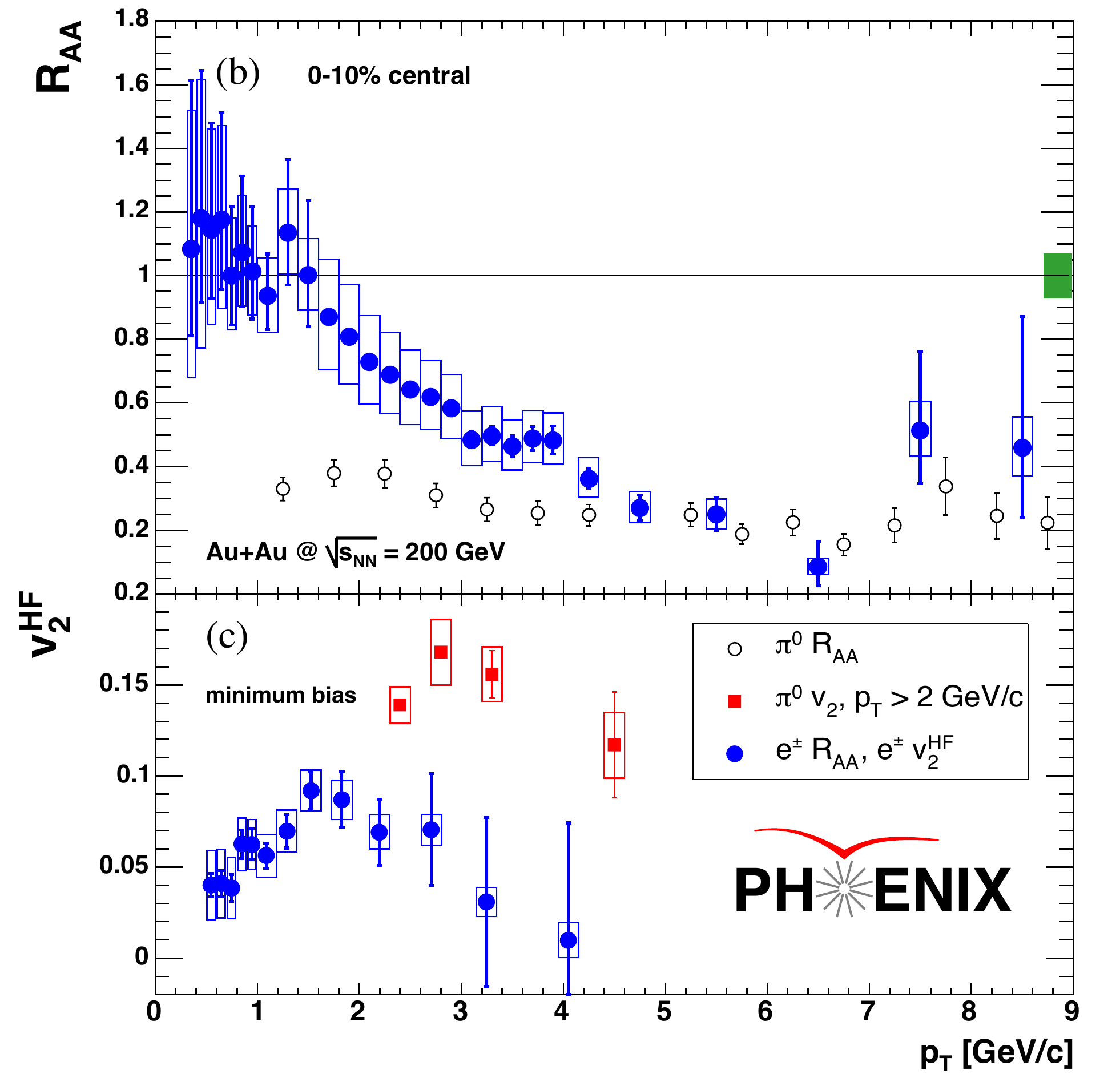}}
\end{center}\vspace*{-2.0pc}
\caption[]{\footnotesize (a) (left) PHENIX~\cite{ppg003} discovery of suppression ($R_{AA}\sim 0.3$) of charged hadrons and $\pi^0$ in Au$+$Au central collisions at \sqsn=130 GeV compared to enhancement ($R_{AA}>1.0)$ at lower \sqsn.\\ (right) PHENIX direct-single-$e^{\pm}$ in Au$+$Au at \sqsn=200 GeV~\cite{ppg066}: (b) $R_{AA}$ (central) (c) $v_2$ (minimum bias) vs. $p_T$, compared to $\pi^0$. }\vspace*{-2.0pc}
\label{fig:suppression}
\end{figure}

   The discovery at RHIC that $\pi^0$'s produced at large transverse momenta are suppressed in central Au$+$Au collisions by a factor of $\sim5$ compared to binary-collision scaling from p$+$p collisions (Fig.~\ref{fig:suppression}a)~\cite{ppg003} is arguably {\em the}  major discovery in Relativistic Heavy Ion Physics. At RHIC energies, the steeply falling exponential spectrum at low $p_T$ becomes negligible relative to the power-law dependence of hard-scattering for $p_T\geq 3$ GeV/c~\cite{ppg063PRD76}. The suppression in A$+$A collisions is presented as $R_{AA}(p_T)$, the ratio of the yield of e.g. $\pi^0$ per A+A collision of a given centrality  to the binary-scaled p$+$p cross section at the same $p_T$, where $\mean{T_{AA}}$ is the average overlap integral of the nuclear thickness functions for that centrality\vspace*{-0.4pc}  
   \begin{equation}
R_{AA}(p_T)=\frac{(1/N_{AA})\;{d^2N^{\pi}_{AA}/dp_T dy}} { \mean{T_{AA}}\;\, d^2\sigma^{\pi}_{pp}/dp_T dy} \quad .
  \label{eq:RAA}\vspace*{-0.4pc}
 \end{equation}
Fig.~\ref{fig:suppression}a illustrates another advantage of high $p_T$ suppression as a \QGP\ probe: at  \sqsn\ of  CERN SpS fixed target and ISR A+A (i.e. $\lsim 31$ GeV), high $p_T$ production is enhanced.

A related but possibly greater discovery, the opposite of what was predicted~\cite{deadcone}, is that heavy quarks observed using direct-single-$e^{\pm}$ from  $c$ and $b$-quark decay are suppressed comparably to light quarks ($\pi^0$) for $p_T\gsim 4$ GeV/c (Fig~\ref{fig:suppression}b) and also exhibit flow (Fig~\ref{fig:suppression}c)~\cite{ppg066}.  The important impact of this discovery was that it provided a demonstration that heavy quarks were strongly coupled to the medium with viscosity/entropy density \mbox{$\eta$/$s\approx (1.3-2)$/$4\pi$}, close to the quantum lower bound~\cite{Kovtun05}, reinforcing the `perfect fluid', and stimulating a broad spectrum of possible explanations (e.g see Ref.~\cite{MJTIJMPA2014}).  
\section{RHIC Beam Energy Scan (BES) in search of a critical point---aided by  Lattice \QCD}
This subject has remained one of my main interests since I got sandbagged by a press release from LBL during ISSP2011~\cite{MJTIJMPA2014} claiming that ``By comparing theory with data from STAR, Berkeley Lab scientists and their colleagues map phase changes in the \QGP" after I had criticized a different laboratory for physics by press release. It turned out that the LBL press release and publication were later found to be ``not useful''~\cite{KarschRedlichPRD84} (i.e. ``not correct''~\cite{MJTIJMPA2014}). The good news is that this year my colleagues and I at PHENIX have actually made such a comparison and were able to use our measurements of net-charge fluctuations~\cite{ppg179} together with a Lattice \QCD\ calculation~\cite{BNLlattice2012} to find both the freezeout temperature, $T_f$, and the Baryon Chemical Potential $\mu_B$ at several values of \sqsn at RHIC, without particle identification!

\begin{figure}[!h]
   \begin{center}
\raisebox{0.2pc}{\includegraphics[width=0.38\textwidth]{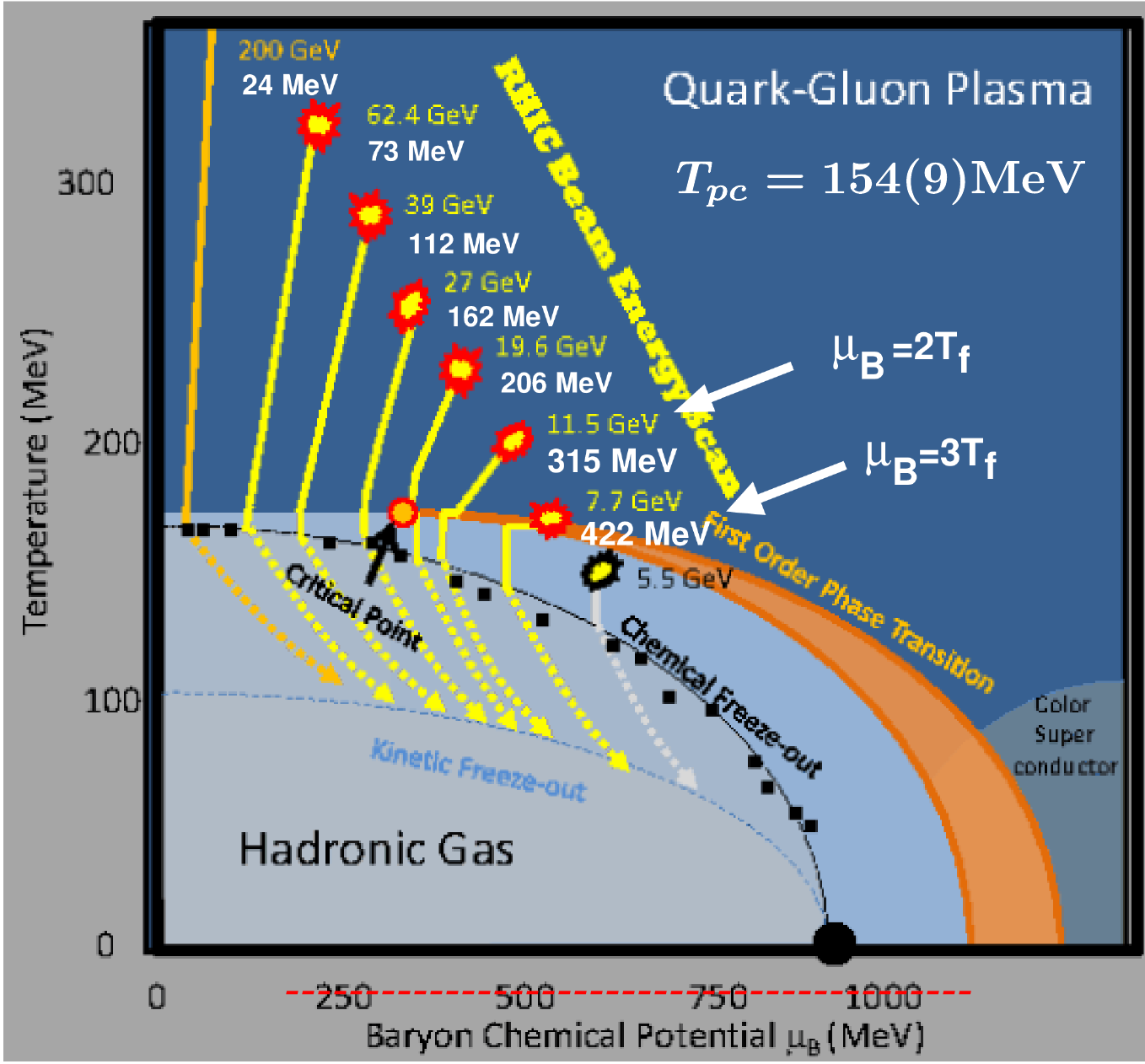}}\hspace{1pc}
\raisebox{0.0pc}{\includegraphics[width=0.42\textwidth]{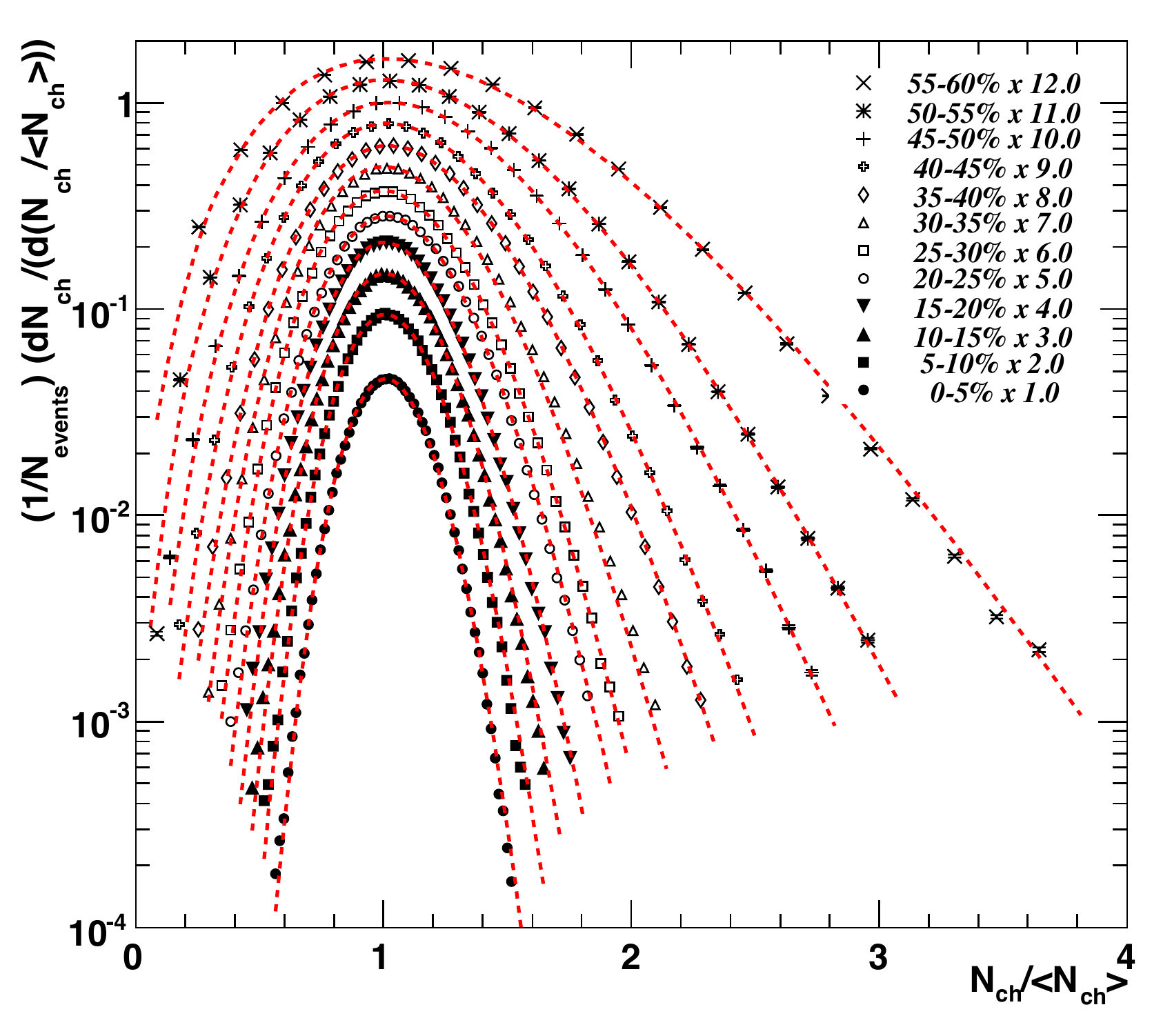}}
\end{center}\vspace*{-1.0pc}
\caption[]{\footnotesize (a) Figure 20a from Ref.~\cite{MJTISSP2014Proceedings} corrected by F.~Karsch. (b) PHENIX~\cite{PXPRC78} fits of Negative Binomial Distributions to $\Nch/\mean{\Nch}$ measurements from \sqsn=200 GeV Au$+$Au for the centralities indicated.      }
\label{fig:PhaseDiagramNBD}
\end{figure}

To set the stage for the new measurement, Fig.~\ref{fig:PhaseDiagramNBD}a shows a proposed phase diagram for nuclear matter (which I emphasized in last year's proceedings~\cite{MJTISSP2014Proceedings} had an incorrect $\mu_B$ scale) as corrected by Frithjof Karsch; and Fig.~\ref{fig:PhaseDiagramNBD}b shows measurements of the total charge multiplicity $\Nch=\Nch^+ + \Nch^-$ as a function of centrality in Au+Au collisions at RHIC together with fits to Negative Binomial Distributions (NBD)~\cite{PXPRC78}, \fbox{$P(m)=\frac{(m+k-1)!}{m!(k-1)!}\frac{({\mu}/{k})^m}{(1+{\mu}/{k})^{m+k}}$}, where $P(m)$ is normalized from $0\leq m\leq\infty$ and $\mu\equiv\mean{\Nch}=\mean{m}$. 

\begin{figure}[!h]\vspace*{-1.0pc}
   \begin{center}
\raisebox{0.0pc}{\includegraphics[width=0.75\textwidth]{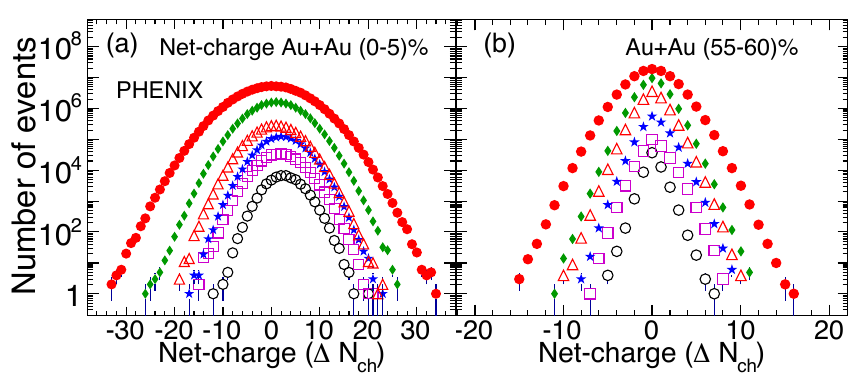}}
\end{center}\vspace*{-2.0pc}
\caption[]{\footnotesize PHENIX~\cite{ppg179} $\Delta\Nch=\Nch^+ - \Nch^-$ distributions uncorrected for the efficiency within the acceptance, $|\eta|<0.35$, $\delta\phi=\pi$, $0.3<p_T<2.0$ GeV/c, in Au$+$Au at two centralities, for all \sqsn measured.  }\vspace*{-0.5pc}
\label{fig:PXnetchraw}
\end{figure}

Figure~\ref{fig:PXnetchraw} shows the new PHENIX $\Delta\Nch=\Nch^+ - \Nch^-$ distributions from Au$+$Au collisions, not yet corrected for the reconstruction efficiency $\epsilon\approx 0.69$ within the acceptance~\cite{ppg179}. 
These distributions can not be compared directly to Lattice \QCD\ predictions which use cumulants from a Taylor expansion of the free energy $F=-T\ln Z$ around the critical temperature $T_c$, where $Z$ is the partition function (or sum over states) which is of the form $Z\propto e^{-(E-\sum_i \mu_i Q_i)/kT}$ and $\mu_i$ are chemical potentials associated with conserved charges.  The terms of the Taylor expansion, which are obtained by differentiation, are called susceptibilities, denoted $\chi$. The connection of this method to the measurements is that the cumulant generating function in mathematical statistics is also a Taylor expansion of the $\ln$ of an exponential, where the $n$-th cumulant $\kappa_n$ represents the $n$-th moment of a distribution with all $n$-fold combinations of the lower order moments subtracted. 
One of the most important properties of cumulants is that they are additive for the sum of statistically independent random variables, $x$ and $y$ where $\mean{xy}=\mean{x}\mean{y}$:  \mbox{$\kappa_n(x+y)=\kappa_n(x)+\kappa_n(y)$}. 

A recent important theorem from (of all places) Quantitative Finance~\cite{Barndorff2013} 
for  ``integer value Levy processes'' (notably Poisson and Negative Binomial Distributions) has proved that the cumulants $\kappa_j$ for the distribution $P(x-y)$ of the difference of samples from two such distributions, $P^{+}(x)$ and $P^{-}(y)$, with cumulants $\kappa_j^+$ and $\kappa_j^-$, respectively, are \fbox{$\kappa_j=\kappa_j^+ +(-1)^j \kappa_j^-$} so long as the distributions are not 100\% correlated. The individual $\Nch^+$ and $\Nch^-$ distributions are in fact NBD~\cite{ppg179} so this theorem would allow a great simplification of the analysis but is presently used only as a check of the very complicated standard method which starts with the calculation of the cumulants of the  $\Delta\Nch=\Nch^+ - \Nch^-$ distributions from Fig.~\ref{fig:PXnetchraw}. 

\begin{figure}[!bh]\vspace*{-1pc}
   \begin{center}
\raisebox{0.15pc}{\includegraphics[width=0.345\textwidth]{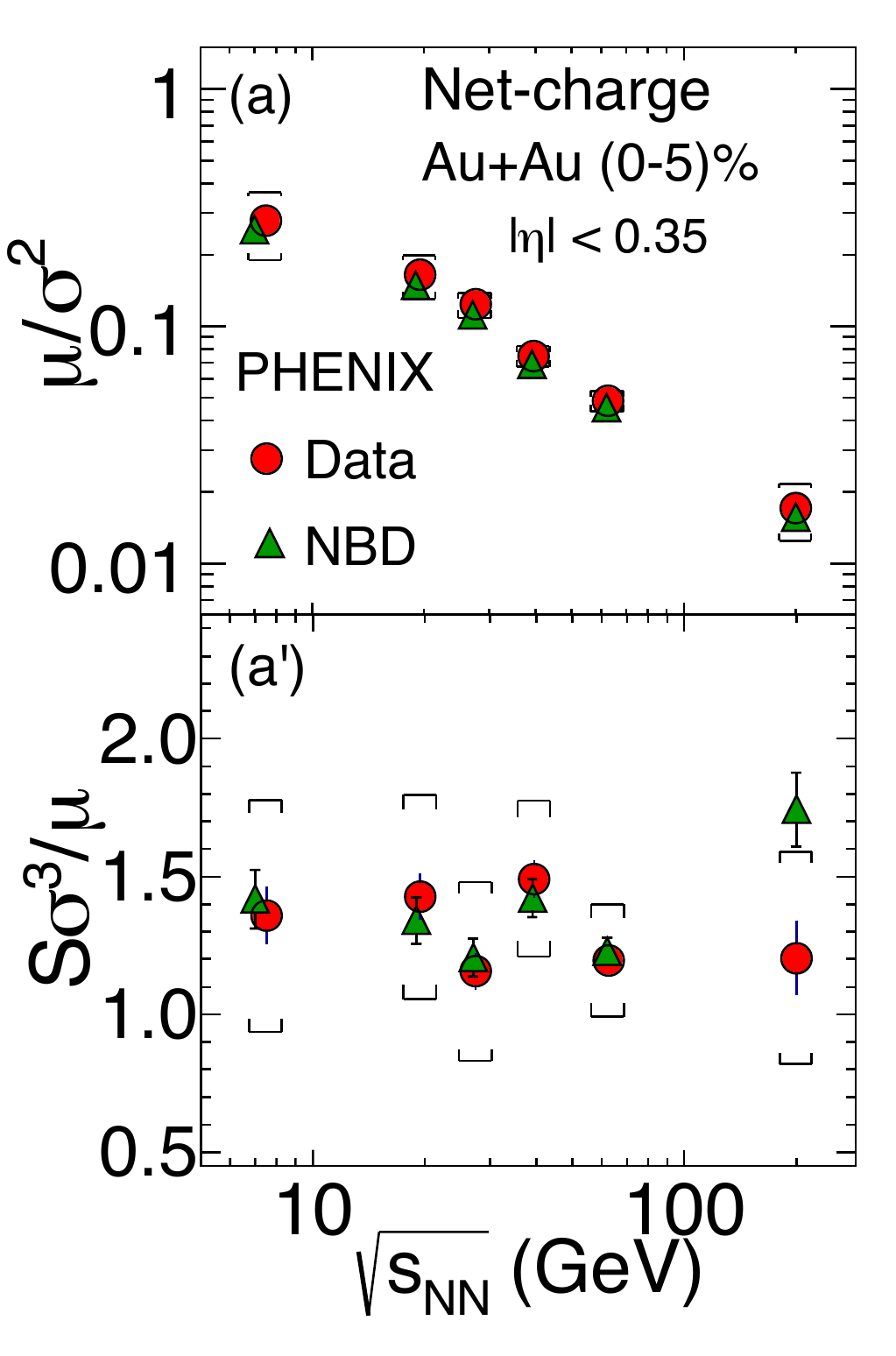}}
\raisebox{0.60pc}{\includegraphics[width=0.325\textwidth]{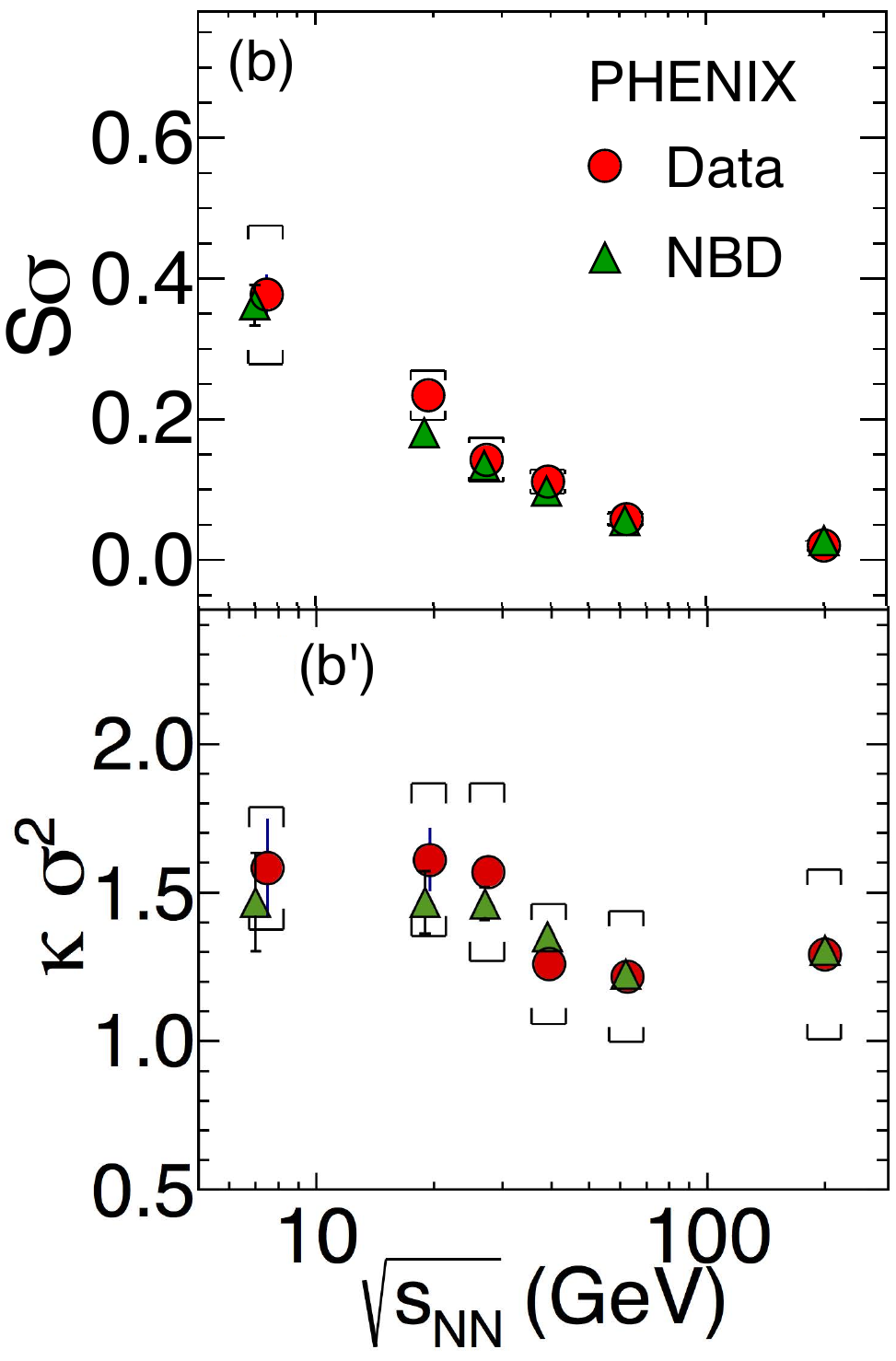}}
\raisebox{0.0pc}{\includegraphics[width=0.315\textwidth,height=0.505\textwidth]{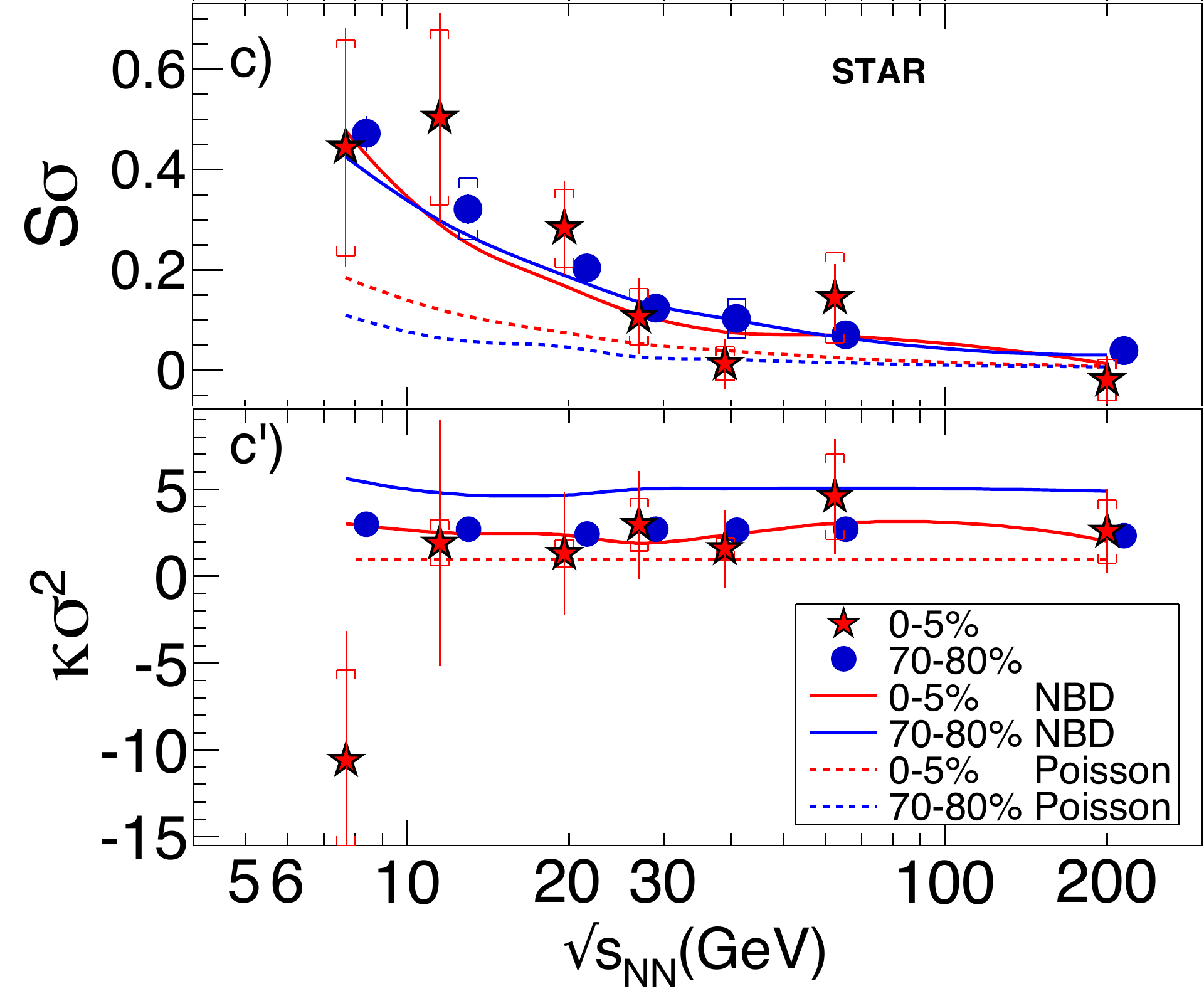}}
\end{center}\vspace*{-1.5pc}
\caption[]{\footnotesize \sqsn dependence of net-charge cumulant ratios in Au$+$Au central (0-5\%) collisions. PHENIX~\cite{ppg179}: (a) $\kappa_1/\kappa_2$, (a$'$) $\kappa_3/\kappa_1$, (b) $\kappa_3/\kappa_2$, (b$'$) $\kappa_4/\kappa_2$. STAR~\cite{STARNetCh}: (c) $\kappa_3/\kappa_2$, (c$'$) $\kappa_4/\kappa_2$. The error bars are statistical and the caps are systematic uncertainties. }\vspace*{-0.0pc}
\label{fig:PXSTAR}
\end{figure}

For each distribution the $n$-th order moment can be calculated: $\mu'_n\equiv \mean{x^n}=\sum_i 
 x_i^n P(x_i)$, where $\mu'_1\equiv \mu=\mean{x}$, $\mu_n\equiv\mean{(x-\mu)^n}$ and $x_i$ represents a bin in the $\Delta\Nch$ plot with a fraction $P(x_i)$ of the total number of events. Then one computes the cumulants $\kappa_m$ from the moments, where presently the analyses compute only the first four  cumulants: \vspace*{-1.0pc}
 \begin{multline}
{\color{PineGreen}\mu=}\kappa_1=\mu^{'}_1 \qquad
{\color{PineGreen}\sigma^2=\mu_2\equiv\mean{(x-\mu)^2}=}\kappa_2=\mu^{'}_2 -{\mu^{'}_1}^{2}\qquad
{\color{PineGreen}\mu_3=}\kappa_3=\mu^{'}_3 -3{\mu^{'}_2}{\mu^{'}_1}+2{\mu^{'}_1}^{3}\\
{\color{PineGreen}\mu_4-3\mu_2^2=}\kappa_4={\mu^{'}_4}-4{\mu^{'}_3}{\mu^{'}_1}-3{\mu^{'}_2}^{2}+12{\mu^{'}_2}{\mu^{'}_1}^{2}-6{\mu^{'}_1}^{4}\qquad\quad\label{eq:calccumulants}\vspace*{-1.0pc}
\end{multline}
\vspace*{-0.2pc}
The cumulants are then corrected for reconstruction efficiency.
Ratios of the corrected cumulants are used to compare to the calculated ratios of susceptibilities because the volume dependences of the calculations cancel: $\mu/\sigma^2=\kappa_1/\kappa_2\approx \chi_1/\chi_2$; $S\sigma^3/\mu=\kappa_3/\kappa_1\approx\chi_3/\chi_1$; $S\sigma=\kappa_3/\kappa_2\approx \chi_3/\chi_2$; $\kappa\sigma^2=\kappa_4/\kappa_2\approx\chi_4/\chi_2$; where $S\equiv\kappa_3/\sigma^3$ and $\kappa\equiv\kappa_4/\kappa_2^2$ (Fig.~\ref{fig:PXSTAR}).

The new PHENIX net-charge results~\cite{ppg179} for the corrected cumulant ratios shown in Fig.~\ref{fig:PXSTAR}a,b are compared to the previous STAR net-charge measurements~\cite{STARNetCh} for $S\sigma$ and $\kappa\sigma^2$ in Fig.~\ref{fig:PXSTAR}c.  
There are two important observations from this comparison: i) for PHENIX the error on all the corrected cumulant ratios is 20-30\%, while for STAR the error on \mbox{$\kappa_1/\kappa_2=\mu/\sigma^2$} is $<1$\% but for $\kappa_3/\kappa_2=S\sigma$ is $\sim 40$\% and for $\kappa_4/\kappa_2=\kappa\sigma^2$ is $>100$\%; ii) the PHENIX standard cumulant ratios and errors (circles) are checked by NBD fits to the $\Nch^+$ and $\Nch^-$ distributions plus the cumulant theorem~\cite{Barndorff2013} (triangles) and agree---which proves that the errors are correct and that the cumulant theorem works. 

\begin{figure}[!h]
   \begin{center}
\raisebox{0.0pc}{\includegraphics[width=0.40\textwidth]{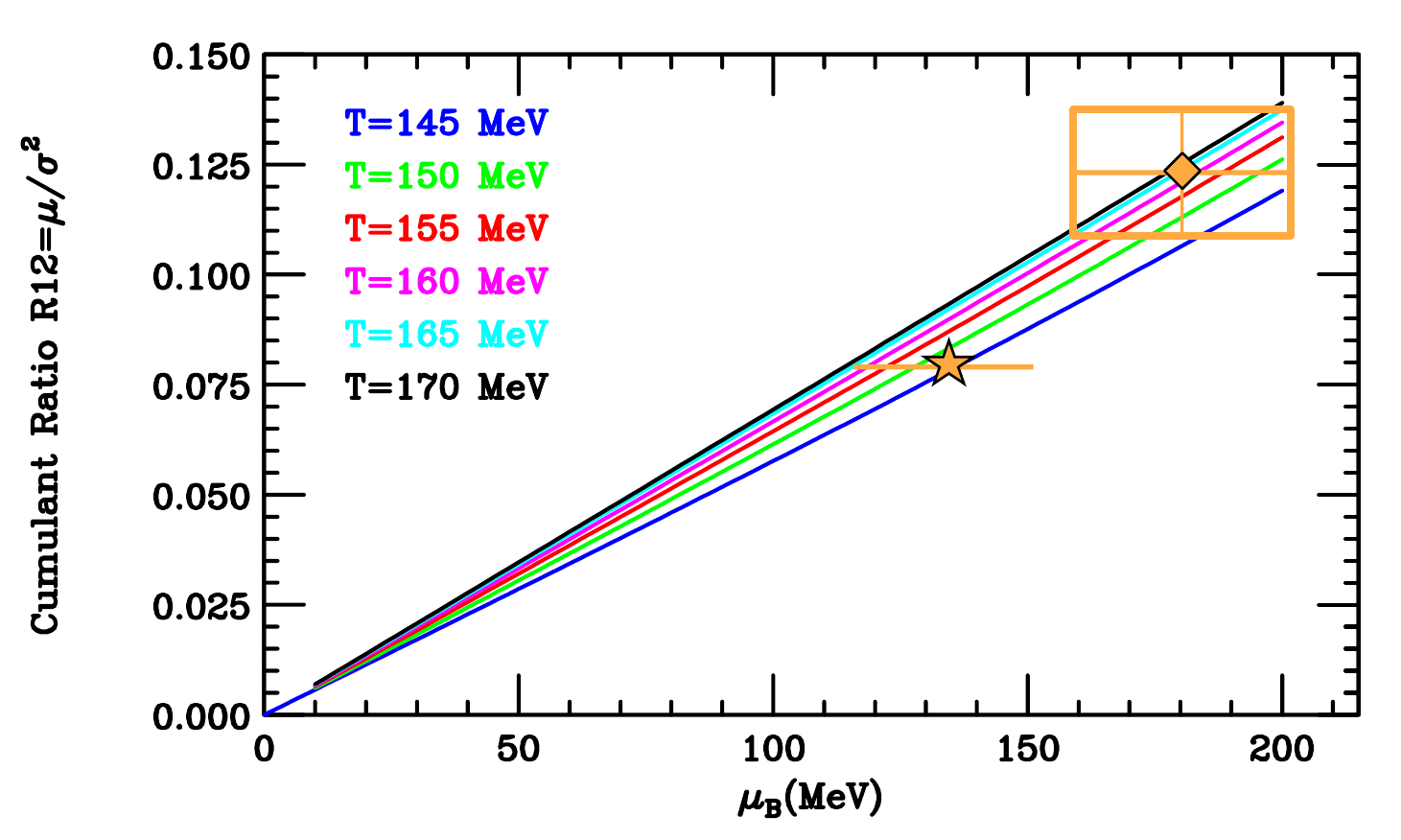}}\hspace*{2.0pc}
\raisebox{0.0pc}{\includegraphics[width=0.40\textwidth]{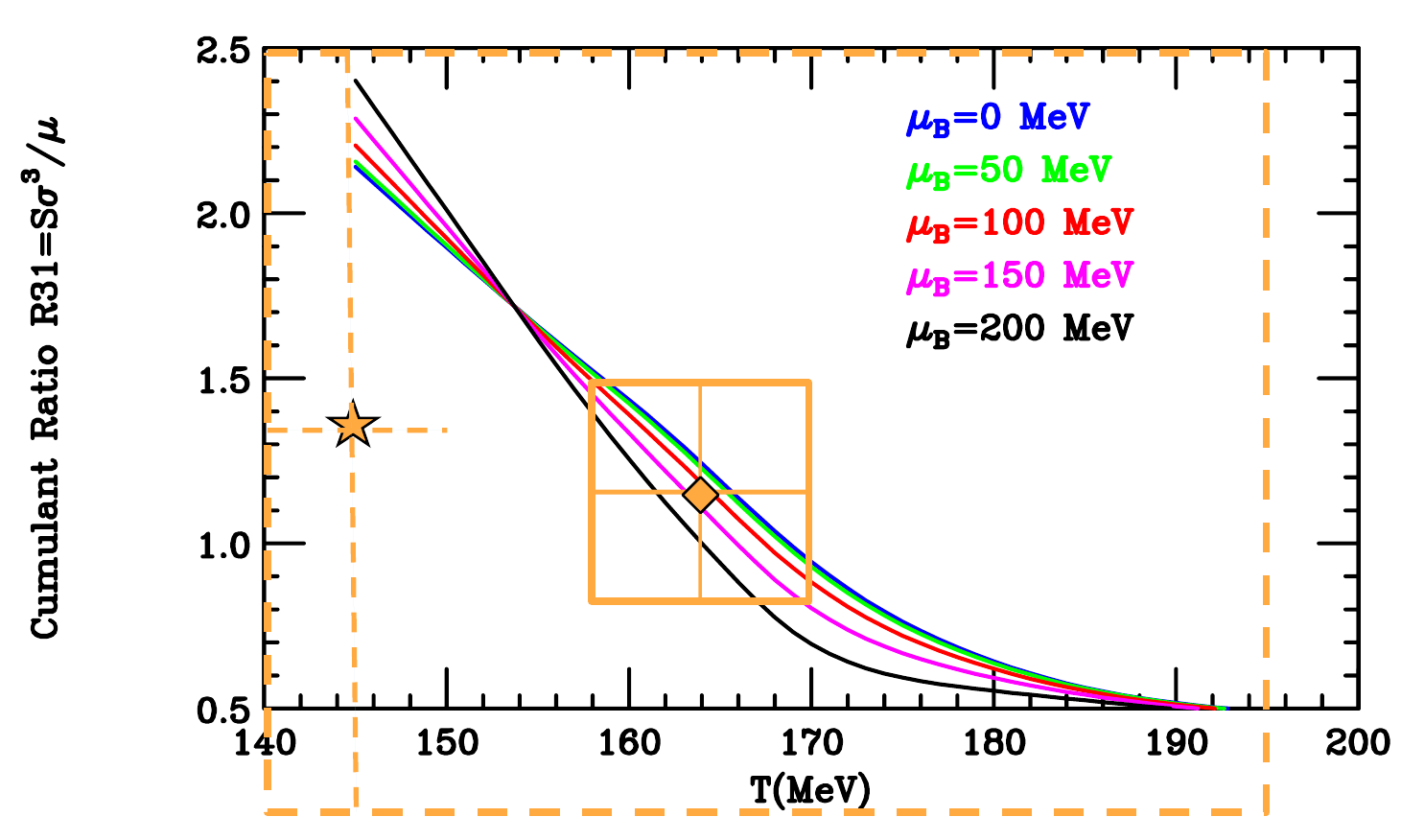}}
\end{center}\vspace*{-1.0pc}
\caption[]{\footnotesize Lattice \QCD\ calculations of cumulant ratios (lines)~\cite{BNLlattice2012} (a) $R_{12}=\kappa_1/\kappa_2$ (b) $R_{31}=\kappa_3/\kappa_1$. The vertical lines in the boxes represent the error of the $R_{12}$ and $R_{31}$ measurements at \sqsn=27 GeV and the horizontal lines the error of the best fit values of $T_f$ and $\mu_B$. The STAR  measurement of $R_{31}$ has a such a huge error that the data point ($\bigstar$) could go anywhere in the dashed box. The horizontal dashes and position of the $\bigstar$ for $R_{31}$ are an assumption~\cite{BorsanyiPRL113} based on the STAR net-proton measurement~\cite{STARPRL112NetP}.  }
\label{fig:LatticeQCD}
\end{figure}

As illustrated in Fig.~\ref{fig:LatticeQCD}, the PHENIX ({\footnotesize $\blacklozenge$}) errors allow both $\mu_B$ and $T_f$  to be determined from the Lattice \QCD\ calculations~\cite{BNLlattice2012} while the large STAR ({\footnotesize $\bigstar$}) error on $R_{31}=\kappa_3/\kappa_1$ does not give a solution for $T_f$.

\begin{figure}[!h]
   \begin{center}
{\small a)}\raisebox{0.0pc}{\includegraphics[width=0.46\textwidth]{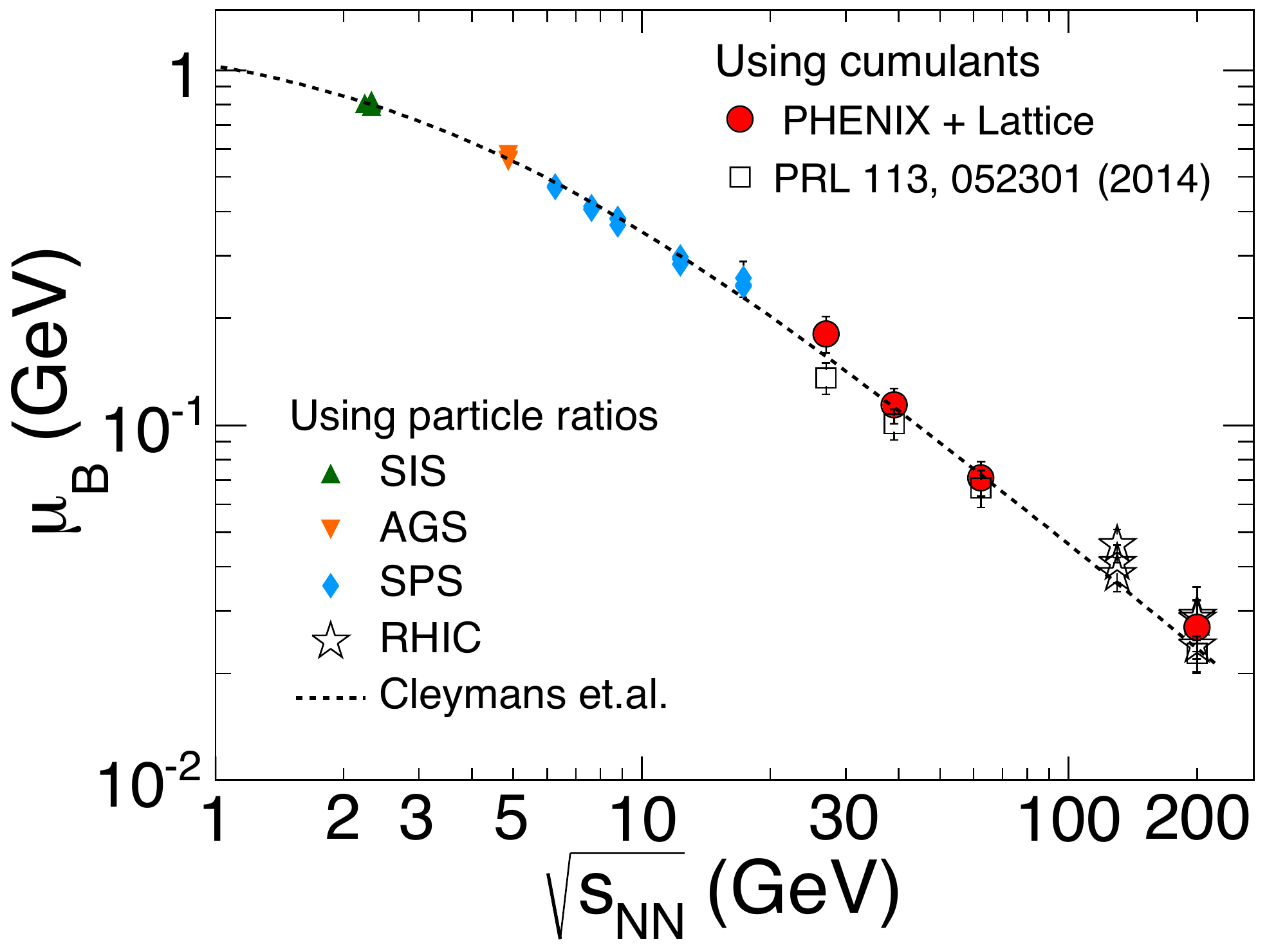}}\hspace*{3.0pc}
{\small b)}\raisebox{1.0pc}{\includegraphics[width=0.29\textwidth]{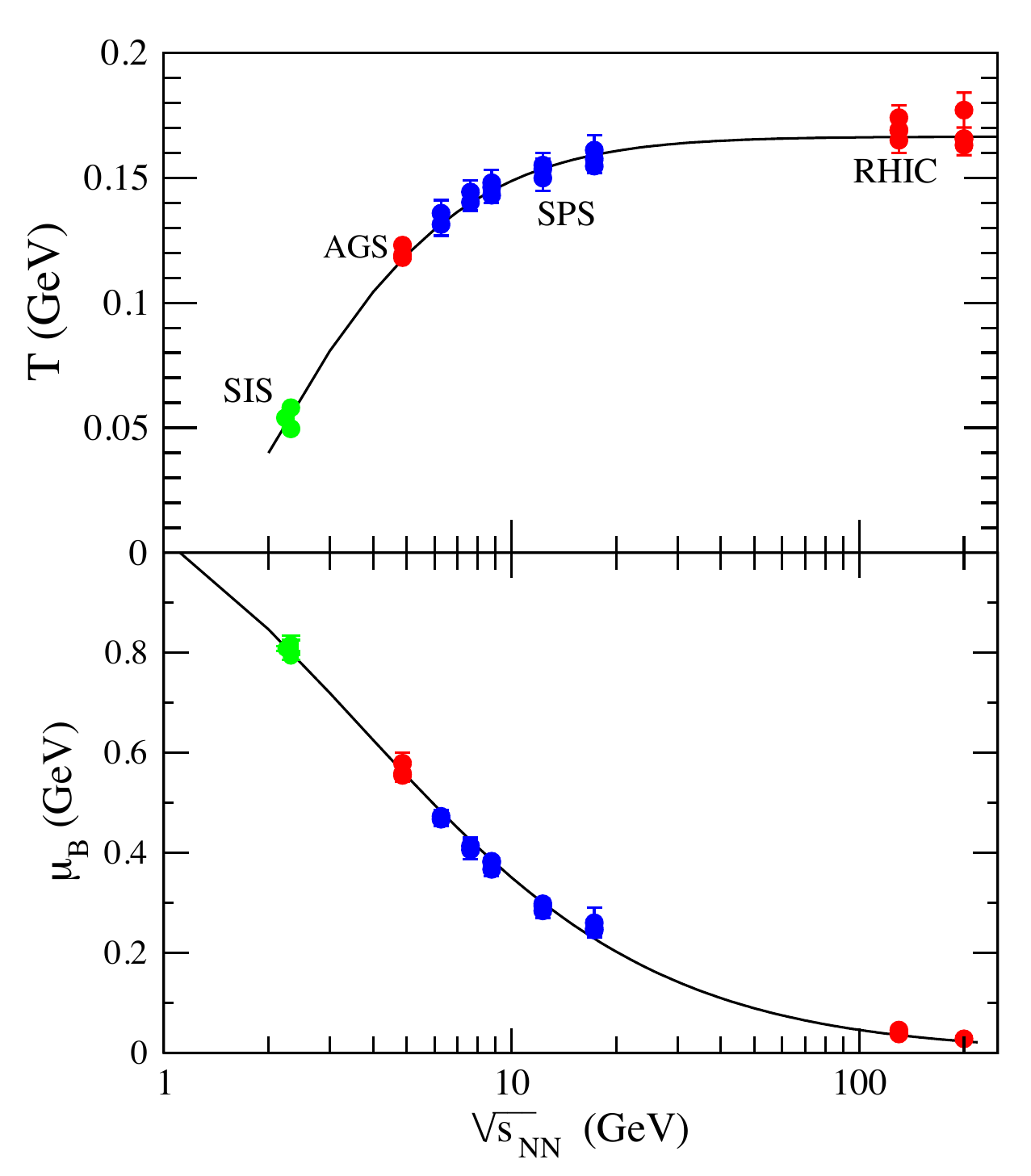}}
\end{center}\vspace*{-1.0pc}
\caption[]{\footnotesize (a) $\sqsn$ dependence of $\mu_B$ from PHENIX+Lattice~\cite{ppg179} net-charge results. Open squares are from STAR net-charge together with net-protons~\cite{BorsanyiPRL113}. Dashed line and other data points are from (b), the best accepted analysis of $T_f$ and $\mu_B$ vs \sqsn from baryon/anti-baryon ratios~\cite{CleymansOeschler}.} \vspace*{-1.0pc}
\label{fig:TmuB}
\end{figure}

Figure~\ref{fig:TmuB} shows that the PHENIX + Lattice results for $T_f$ and $\mu_B$ from net-charge fluctuations, \underline{with no particle identification}, are in excellent agreement with the best accepted analysis of $T_f$ and $\mu_B$ from baryon/anti-baryon ratios.~\cite{CleymansOeschler}. I believe that this is a first! 

\subsection{Why are the STAR errors on $R_{31}$ so large? Corrections?}
A fraction of the tracks that fall on the detector acceptance are not reconstructed, a clearly random, thus Binomial, effect. This is further complicated if the efficiencies are different for $\Nch^+$ ($\epsilon^+$) and $\Nch^-$ ($\epsilon^-$). The standard Koch-Bzdak~\cite{BzdakKoch} Binomial efficiency correction, e.g. to correct $\kappa_3$ of an arbitrary net-charge distribution, is: \vspace*{-0.6pc} 
\begin{equation}\label{eq:complicated}
\kappa_3=\kappa_1+2\kappa_1^3-F_{03}-3F_{02} +3F_{12}+3F_{20}-3F_{21}+F_{30}-3\kappa_1(N+F_{02}-2F_{11}+F_{20})\vspace*{-0.5pc}
\end{equation}
where the corrected values are $\mean{N^+}=\mean{\Nch^+}/\epsilon^+$, $\mean{N^-}=\mean{\Nch^-}/\epsilon^-$, $\kappa_1=\mean{N^+}-\mean{N^-}$, $N\equiv\mean{N^+}+\mean{N^-}$. Also, a new quantity, the double Factorial Moment has been introduced:
\fbox{$F_{ik}=\sum_{N_1=i}^\infty \sum_{N_2=k}^\infty P(N_1,N_2)\frac{N_1!}{(N_1-i)!} \frac{N_2!}{(N_2-k)!}$}, which 
looks simple, but is very difficult to compute because the joint probability $P(N_1,N_2)$, e.g. 
$P(13^+, 11^-)$ etc., is required. This means that to make the computation one must know not only the $\Delta \Nch$ distribution but also both the $N^+$, the $N^-$ distributions and their correlation. $P(N^+,N^-)$ can be obtained from the data by making a 3d lego plot with base axes $N^+$ and $N^-$ and height equal to the normalized number of events with ($N^+$, $N^-$), but this costs a huge dilution of the statistical errors, so approximations or random-sampling of the lego plot are used. My guess is that this dilution plus the huge complication in e.g. the calculation of $\kappa_3$ in Eq.~\ref{eq:complicated} are one reason for the huge STAR errors in $S\sigma=\kappa_3/\kappa_2$ and $\kappa\sigma^2=\kappa_4/\kappa_2$ but I haven't verified it with them.
\subsubsection{If you know the distribution, you know all the moments and cumulants}
When I first saw net-charge distribution plots like Fig.~\ref{fig:PXnetchraw}, e.g. see Ref.~\cite{MJTIJMPA2014}, my thought was to fit them to NBD so that the title of this section would apply. However, they are not NBD, they are the difference between two NBD. This is why the cumulant theorem~\cite{Barndorff2013} is so important: once the cumulants, $\kappa_j^+$ and $\kappa_j^-$, of NBD fits to the individual $\Nch^+$ and $\Nch^-$ are known, the cumulants of the $\Delta\Nch$ distribution are simply {$\kappa_j=\kappa_j^+ +(-1)^j \kappa_j^-$}. 
Perhaps equally important, the correction for efficiency to the measured $\Nch^+$ and $\Nch^-$ is a ``binomial split'' of an NBD~\cite{CarruthersShih1985}---if a population is NBD$(\mu_t,k)$ and divided randomly with probability $p$ onto a smaller region, the distribution on the region is NBD$(p\mu_t,k)$, i.e. the mean changes but the $k$ parameter remains constant. Thus if we measure $\mu^+=\mean{\Nch^+}$ in the detector with efficiency $p=\epsilon^+$, the efficiency corrected mean,  $\mu_t^+$ equals $\mu^+/\epsilon^+$ and e.g. the efficiency corrected $\kappa_3/\kappa_1$ is: \vspace*{-0.6pc} 
\begin{equation}
\frac{S\sigma^3}{\mu}=\frac{\kappa_3}{\kappa_1}=\frac{\kappa_3^+ -\kappa_3^-}{\kappa_1^+ -\kappa_1^-}
=\frac{\mu_t^+ [1+3(\frac{\mu_t^+}{k^+})+2(\frac{\mu_t^+}{k^+})^2]-\mu_t^- [1+3(\frac{\mu_t^-}{k^-})+2(\frac{\mu_t^-}{k^-})^2]}{\mu_t^+ -\mu_t^-}  \qquad ,
\label{eq:Ssig3overmu}\vspace*{-0.5pc}
\end{equation}    
which is a lot simpler than Eq.~\ref{eq:complicated} and dramatically simpler than for $\kappa_4/\kappa_2=\kappa\sigma^2$, which the reader can figure out from Table~\ref{tab:CCumulant} compared to Ref.~\cite{BzdakKoch}. 

 \begin{table}[!ht]\vspace*{-1.0pc}
\begin{center}
\caption[]{Cumulants for Poisson, Negative Binomial Distributions\\\hspace*{4pc}Measured with efficiency $\epsilon$ corrected to the true value}
{\begin{tabular}{llll} 
\hline
Measured Cumulant & Corrected Poisson & Corrected Negative Binomial \\
\hline
$\kappa_1=\mu$ & $\mu_t\equiv\mu/\epsilon$ &$\mu_t\equiv\mu/\epsilon$\\
$\kappa_2=\mu_2=\sigma^2$ &$\mu_t$&  $\mu_t (1+{\mu_t}/{k})\equiv\sigma_t^2$\\[0.2pc]
$\kappa_3=\mu_3$ &$\mu_t$& $\sigma_t^2 (1+2{\mu_t}/{k})$\\[0.2pc] 
$\kappa_4=\mu_4-3\kappa_2^2$\hspace*{1pc}&$\mu_t$&  $\sigma_t^2 (1+6{\mu_t}/{k}+6{\mu_t^2}/{k^2})$\\[0.2pc]
\hline\\[-0.6pc]
$S\equiv{\kappa_3}/{\sigma^3}$& $ {1}/{\sqrt{\mu_t}}$  & $(1+2{\mu_t}/{k})/{\sigma_t}$\\[0.2pc]
$\kappa\equiv{\kappa_4}/{\kappa_2^2}$ & ${1}/{\mu_t}$ &  $(1+6{\mu_t}/{k}+6{\mu_t^2}/{k^2})/{\sigma_t^2}$\\[0.2pc]
$S\sigma=\kappa_3/\kappa_2$ & $1$ &  $(1+2{\mu_t}/{k})$\\[0.2pc]
$\kappa\sigma^2=\kappa_4/\kappa_2$ & $1$ &  $(1+6{\mu_t}/{k}+6{\mu_t^2}/{k^2})$\\[0.2pc]
$\mu/\sigma^2=\kappa_1/\kappa_2$ & $1$ & $1/(1+{\mu_t}/{k})$\\[0.2pc] 
$S\sigma^3/\mu=\kappa_3/\kappa_1$ & $1$ & $(1+3{\mu_t}/{k}+2{\mu_t^2}/{k^2})$\\[0.2pc] 
\hline
\end{tabular}} \label{tab:CCumulant}
\end{center}\vspace*{-2pc}
\end{table}

\subsection{The End?}Of course this is still not the whole story.
Since there can be no fluctuations of conserved quantities such as net-charge or net-baryon number in the full phase space, one must go to  ``locally conserved quantities''~\cite{AsakawaHMPRL85} in small rapidity intervals to detect a fraction which then fluctuates, i.e. varies from event to event. Then the question becomes what the required acceptance is for an experimental result to compare with the Lattice \QCD\ calculations.  Another issue is whether ``net-protons'' are equivalent to net-baryons since the neutrons are not measured. Further complicating this, the experiments measure in a fixed $\delta\eta$ interval but the rapidity distributions change (dramatically for protons because of stopping~\cite{BRAHMSNetPPRL93}) as the \sqsn is reduced. Also Lattice \QCD\ calculations are not now available for $\mu_B> 0.2$ GeV\ldots There are many unresolved isues---stay tuned!!!

\end{document}